\documentclass[openacc]{rsos}

\usepackage{orcidlink}

\newtheorem{theorem}{\bf Theorem}[section]

\newtheorem{corollary}{\bf Corollary}[section]

\newtheorem{algorithm}{\bf Algorithm}[section]
\newtheorem{definition}{\bf Definition}[section]
\newtheorem{example}{\bf Example}[section]

\titlehead{Research}

\begin{document}
	
	\title{Transition Graphs of Interacting Hysterons: Structure, Design, Organization and
		Statistics}
	
	\author{
		Margot H. Teunisse\orcidlink{0009-0005-2289-6148}$^{1, 2}$ and Martin van Hecke\orcidlink{0000-0002-9550-6607}$^{1, 2}$}
	
	\address{$^{1}$Huygens-Kamerlingh Onnes Lab, Universiteit Leiden, P.O.~Box~9504, 2300 RA, Leiden, Netherlands\\
		$^{2}$AMOLF, Science Park 104, 1098 XG, Amsterdam, Netherlands\\}
	
	\subject{statistical physics, combinatorics}
	
\corres{Margot Teunisse\\
	\email{teunisse@physics.leidenuniv.nl}\\
	Martin van Hecke\\
	\email{M.v.Hecke@amolf.nl}}

\begin{abstract}
	Transition graphs capture the memory and sequential response of multistable media, by specifying their evolution under external driving. Microscopically, collections of bistable elements, or hysterons, provide a powerful model for these materials, with recent work highlighting the crucial role of hysteron interactions. Here we introduce a general framework that links transition graphs and the microscopic parameters of interacting hysterons. We first introduce a systematic framework, based on so-called scaffolds, which structures the space of transition graphs, and provides tools to deal with their combinatorial explosion. We then connect the topology of transition graphs to partial orders of the microscopic parameters. This allows us to understand the statistical properties of transition graphs, as well as determine whether a given graph is realizable, i.e., compatible with the hysteron framework. Our approach paves the way for a deeper theoretical understanding of memory effects in complex media, and opens a route to rationally design pathways and memory effects in materials.
\end{abstract}


\begin{fmtext}
	Multistable systems, ranging from metamaterials to crumpled sheets, exhibit complex responses to sequential driving, often encoding memories of past inputs \cite{keim2019memory}. For example, the sequential response of a crumpled sheet stores intricate memories within its configuration \cite{shohat2021}.
	Moreover, memory effects play a central role in novel physics-based computing approaches
	\cite{liu2024,markovic2020neuromorphic}.
\end{fmtext}

\maketitle

\begin{figure}
	\includegraphics[width=.8\textwidth]{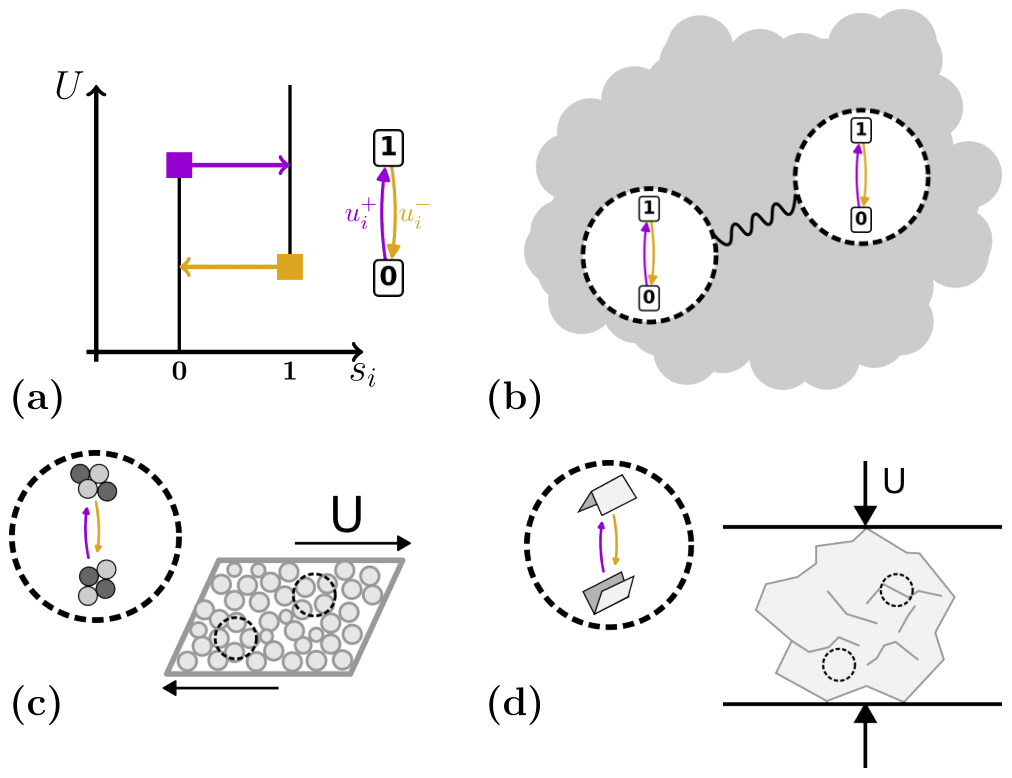}
	\caption{{Interacting hysterons and the physical systems they inhabit. a) The response to driving with a field $U$ (left) and graph representation (right) of a single hysteron with switching fields $u_i^\pm$. The phase of the hysteron is expressed in a binary variable $s_i$. b) Abstract representation of hysterons interacting through their embedding medium. c) Interacting hysterons (local rearrangements) in a sheared amorphous solid \cite{regev2019}. d) Interacting hysterons (mountain/valley folds) in a compressed crumpled sheet \cite{shohat2022memory}.}}
	\label{fig:introduction}
\end{figure}

For {athermal systems that are driven quasistatically}, both the multistability and sequential response can be encoded in a transition graph (t-graph) \cite{keimpaulsen2019,regev2019, mungan2019structure, vhecke2021}.  These represent all stable states as nodes,  connected by directed edges which represent transitions that occur when the driving $U$ exceeds specific critical values.
The beauty of these t-graphs is that they encode the response to {\em any} driving protocol, thus providing a central tool to study the properties and memory effects \cite{paulsen2024mechanical, lindeman2025generalizing} of {a variety of} frustrated media {including amorphous solids} \cite{regev2019, lindeman2021,keimpaulsen2021,szulc2022cooperative}, sequential biological evolution \cite{das2022}, crumpled thin sheets \cite{shohat2021,shohat2024crumpled,shohat2025aging}, corrugated sheets \cite{bense2021} and metamaterials \cite{juleslechenault2021, merrigan2022disorder, liu2024, el2024tunable,muhaxheri2024bifurcations,melancon2022,ducarme2025exotic,shohat2025geometric,paulsen2024mechanical,muhaxheri2025catastrophic}.

To link t-graphs to the microscopic properties of physical models, simulations of frustrated media evolving in complex energy landscapes can be used \cite{regev2019,regev2021topology,lindeman2023competition}. {Such simulations are computationally expensive, as they consider all degrees of freedom of the entire system. In many systems, however, the evolution is dominated by a small amount of  irreversible events associated with localized elements (Fig.~\ref{fig:introduction}).
	For example, rearrangements in amorphous particle packings are often composed of rearrangements or 'T1' events of localized clusters of particles. These can be described as hysteretic, two phase elements, that are refered to as hysterons (Fig.~\ref{fig:introduction}a).
	When sheared, the system's evolution is a mix of smooth episodes, interpunctured by sharp and irreversible rearrangements where these localized clusters change configuration \cite{regev2019}. Each of such changes corresponds to one or more hysterons changing their binary phase (Fig.~\ref{fig:introduction}a).
	Thus, a computationally efficient method to study memory {in such systems} is 
	to directly investigate the t-graphs exhibited by  collections of {hysterons} \cite{mungan2019structure,terzimungan2020,vhecke2021,bense2021,lindeman2021,shohat2021,keimpaulsen2021,regev2021topology,juleslechenault2021,das2022,lindemankeim2023,liu2024,szulcmunganregev2022cooperative}.}

{The modelling of multistable systems through hysterons was introduced by Ferenc Preisach, leading to the so-called Preisach model of hysteresis \cite{preisach1935}.
	The model was
	originally formulated in the context of magnetic domains that do not experience (significant) interactions, and thus focused on collections of {\em independent} hysterons.
	The structure and multiplicity of the  t-graphs of the Preisach model have been characterized in detail \cite{mungan2019structure, terzimungan2020}.}

{More recently, the hysteron model has been applied to geometrically frustrated media. However, here the hysterons experience interactions (Fig.~\ref{fig:introduction}b). Going back to our earlier example,
	the clusters in amorphous packings
	are embedded in an elastic matrix, so that each rearrangement influences the other rearrangements
	(Fig.~\ref{fig:introduction}c). Similarly, in other geometrically frustrated media such as 
	crumpled sheets (Fig.~\ref{fig:introduction}d), local 
	changes  can also be described as interacting hysterons 
	\cite{bense2021,shohat2021,ding2022sequential,melancon2022,hyatt2023programming}. }

{Interactions between hysterons are thus emerging as a crucial ingredient to understand the response of many frustrated physical systems \cite{lindeman2021,keimpaulsen2021,vhecke2021,szulcmunganregev2022cooperative,regev2019,bense2021,shohat2021,liu2024}. 
	Such interactions have been directly observed in experiments on macroscopic frustrated materials \cite{bense2021,shohat2021} and metamaterials \cite{liu2024,el2024tunable,muhaxheri2024bifurcations,merrigan2022disorder,melancon2022,hyatt2023programming}.
	In some cases, these interactions can also be calculated from first principles \cite{liu2024,shohat2025geometric,paulsen2024mechanical}. Moreover, under cyclic driving, these materials exhibit complex responses, including long transients and multi-periodic orbits \cite{bense2021,keimpaulsen2021,lindeman2021,lindeman2025generalizing}, that can not be understood in the Preisach model, but that can be captured in models of interacting hysterons. }

Numerical explorations of models of interacting hysterons have shown that even weak interactions lead to a dramatic increase in the number and complexity of the t-graphs \cite{vhecke2021,lindeman2021,keimpaulsen2021}.
For example, while the number of t-graphs in the Preisach model grows with the number of hysterons $n$ as $n!$, numerical sampling reveals a much larger number of t-graphs for interacting hysterons (11 t-graphs for $n=2$, and more than $1.5\times 10^4$ t-graphs for $n=3$). While 
the link between microscopic parameters and Preisach graphs, as well as their statistics, are simple, 
this link is much more complex for 
interacting hysterons, which moreover 
displays statistical weights that vary over many orders of magnitude \cite{vhecke2021,terzimungan2020,lindemankeim2023}.
These t-graphs are also much more varied than those of the Preisach model, featuring, e.g., {\em scrambling} and {\em avalanches}, which together lead to an enormous variety of remarkable phenomena such as transient memories,
multiperiodic responses, and even computational capabilities \cite{keimpaulsen2019, lindeman2021,vhecke2021,bense2021,shohat2021,liu2024}. However, we currently lack a systematic approach to understand 
the multiplicity, statistics and variability of
t-graphs of interacting hysterons.

Here we address the following questions:
\begin{enumerate}
	\item How can we understand the combinatorial explosion of the number of t-graphs and how are they organized?
	\item How can we check if a target t-graph can be realized by a set of interacting hysterons, and if so, how can we determine the hysteron parameters?
	\item {How can we understand the wide variation in statistical weight of t-graphs in design space?}
\end{enumerate}

To tackle these questions we present a systematic framework for the t-graphs and design problem for the most general model of interacting hysterons. As microscopic parameters we take 
the state dependent switching fields\footnote{{We note that in more recent papers the term 'switching thresholds' is used for these.}} $U_i^\pm(S)$; this approach encompasses recently studied cases with more limited,  specific parameterizations of the switching fields \cite{lindeman2021, keimpaulsen2021, regev2021topology, liu2024}. 
We start by considering the forward problem of how a given set of switching fields produces a
t-graph. We introduce the concept of the {\em scaffold} which allows to precisely define
scrambling, as well as the systematic construction of transitions and avalanches.
We further clarify how specific switching fields can produce ill-defined t-graphs (section~\ref{sec:model}).
We then consider the inverse problem: given a (part of a) t-graph, what are the corresponding necessary and sufficient conditions on the switching fields? We present a systematic method for obtaining the set of design inequalities, and discuss {how these correspond to a} partial order {on} the switching fields.  {This} partial order straightforwardly allows to determine if a given target t-graph topology can be realized in the interacting hysteron model, and provides the underlying structure of the design space (section~\ref{sec:graphdesign}).
We subsequently consider the construction and organization of all t-graphs for a given number of hysterons $n$. We discuss how all scaffolds can be generated, and derive a simple expression for their number. {We further} show that all possible transitions for a given scaffold can be organized in finite binary {\em trees}, one for each state and initial transition direction (up or down). Combining scaffolds and trees, we obtain all {\em candidate graphs}, that need to be checked for realizability using their design inequalities. This method allows to systematically label all candidate t-graphs - the complexity lies in checking their realizability. As specific examples, we count and construct all possible t-graphs for $n=2$, all scaffolds for $n=3$ and all t-graphs for $n=3$ that contain one or two short avalanche(s) (section ~\ref{sec:allgraphs}).
We finally discuss the statistical weight of t-graphs in design space. In particular, the number of total orders consistent with a given partial order is a proxy for the volume in design space, thus giving insight in the widely varying statistical weight of distinct t-graphs, as well as the percentage of ill-defined t-graphs \cite{vhecke2021,szulcmunganregev2022cooperative,lindeman2021} (section ~\ref{sec:statweight}).
Together,
the combination of the systematic construction of design inequalities and the focus on the scaffold may facilitate the practical design of (meta)materials that realize targeted
pathways, memory effects and embodied computations.

\section{Model, Transitions and T-Graphs}\label{sec:model}
In this section we present
a general model for interacting hysterons, discuss in detail 
the response 
to changes of the global driving parameter $U$, and 
the resulting transition graphs.
All abstract hysteron models describe interactions via a state dependence of the switching fields, which are the critical values of the driving that lead to flipping of the hysterons,
but 
different physical assumptions lead to
different  functional forms of this state dependence. For example,
one may assume that interactions are
reciprocal (so that the effect of hysteron $i$ on the switching fields of hysteron $j$ equals the effect of hysteron $j$ on the switching fields of hysteron $i$),
that they are pairwise, 
or that they do not affect the hysteron span (so that
the effects of hysteron $j$ on the up and down switching fields of hysteron $i$ are equal)
\cite{lindeman2021,keimpaulsen2021,vhecke2021,szulcmunganregev2022cooperative}. 
However, both in 
experiments and in physical models, reciprocity and span invariance are 
{not respected} \cite{liu2024,bense2021,keimpaulsen2021},
and multi-hysteron interactions generically occur in networks of geometrically coupled hysterons
\cite{shohat2024geometric}.
We therefore
focus on the most general formulation of the state dependency, which naturally encompasses models with restricted forms of interactions (section~\ref{sec:general}).

Once the interactions are specified, we consider transitions between states: events 
where one or more hysterons change phase in response to changes in the driving parameter $U$ (section~\ref{sec:drive}). We first discuss the critical hysterons that initiate such transitions, and use these to define the 
{\em scaffold}, which underpins the structure of all transitions and which allows a precise definition of the important property of {\em scrambling} \cite{bense2021, vhecke2021} (section~\ref{ssec:sourcemap}). We then discuss transitions and in particular avalanches and their relation to the scaffold (section~\ref{ssec:transitions}). Finally,
not all choices of interaction parameters lead to collections of well-defined transitions (section~\ref{ssec:ill}). First, the scenario where multiple hysterons become unstable during an avalanche leads to race conditions that cannot be resolved in the model; second, situations where the model predicts an endless cycle of state transitions can occur \cite{vhecke2021,lindeman2021,szulcmunganregev2022cooperative}. Both problems stem from the hysteron model being a coarse grained model that lacks an underlying energy landscape and dynamics. While
the first situation can be resolved via {the introduction of} an additional dynamical rule - for example, by always flipping the hysteron that is furthest from stability\cite{lindeman2021, keimpaulsen2021} - in this paper we consider both situations as ill-defined(section~\ref{ssec:ill}).
Altogether, our framework presents an unambiguous mapping from a general set of state-dependent switching fields to a t-graph.

\subsection{General Model for Interacting Hysterons}\label{sec:general}
We now detail the general model for a system of $n$ interacting hysterons.
First, let us recall the definition of a hysteron.
Each hysteron $i$ is a bistable element characterized by its binary phase
{$s_i$}---we use the term  phase for individual elements to avoid confusion with 
the collective state.
The two
switching fields $u_i^\pm$ ($u_i^+ > u_i^-$) determine 
the hysteron's response
under driving with an external field $U$. A hysteron in phase $s_i = 0$ is stable when $U < u_i^+$, but when $U > u_i^+$ it is unstable and
switches from 0 to 1 -- the hysteron 'flips up'. Similarly,
a hysteron in phase $s_i = 1$ is stable when $U > u_i^-$, but when $U < u_i^-$ it is unstable and switches from 1 to 0 -- the hysteron 'flips down'.
This response forms an elementary hysteresis loop.

For a collection of $n$ hysterons, we define the collective state $S$ as
\begin{equation}
	S = (s_1, s_2, \dots, s_n)~.
\end{equation}
We denote the saturated states
where all hysterons are either 0 or 1 as $(0 0 \dots)$ and $(1 1 \dots)$.
The set of hysteron indices for which $s_i = 0$ ($s_i = 1$) is referred to as $I_0(S)$ ($I_1(S)$) \cite{das2022}.

We model interactions between hysterons 
via state-dependent hysteron switching fields
$U_i^\pm(S)$:
\begin{equation}\label{eq:general_model}
	U_i^\pm(S) = u_i^\pm + \Delta_i^\pm(S)~,
\end{equation}
where $\Delta_i$ captures the dependence of the switching fields of hysteron $i$
on the collective state $S$.
The switching fields encode when hysterons become unstable:  starting from state $S$, hysteron $i$ becomes unstable when $U$ is increased (decreased) beyond 
$U^+_i(S)$ ($U_i^-(S)$).

We 
assume that the hysteron switching fields $U_i^\pm(S)$ are non-degenerate and finite -- this implies that the saturated states are always
reached when
$U\rightarrow-\infty$ and $U\rightarrow\infty$.
The interaction term $\Delta_i^\pm(S)$ can encode any specific model for hysteron interactions, including the Preisach model (
$\Delta_i^\pm(S) = 0$ \cite{preisach1935}),
{pairwise} interactions ($\Delta_i^\pm = -\Sigma_j c^\pm_{ij}s_j$), reciprocal
interactions 
\cite{keimpaulsen2021}, or an {equal} shift of upper and lower switching fields ($\Delta_i^+(S)=\Delta_i^-(S)$).
We finally note that
switching fields $U_i^+(S)$ ($U_i^-(S)$) are only relevant
when the corresponding hysteron $i$ is in phase 0 (1).
For example, for the state $(001)$, the only relevant switching fields are $U_1^+(001), U_2^+(001)$ and $ U_3^-(001)$ -- in general
there are $n$ switching fields per state.
Hence, while \eqref{eq:general_model} defines $2n\cdot 2^n$ switching fields,
only {half} of these are relevant: $n$ interacting hysterons are defined by precisely $n\cdot 2^n $ switching fields.

\subsection{Response to Driving}\label{sec:drive}

We now consider how a collection of $n$ interacting hysterons responds to driving, i.e., changes in $U$ (Fig.~\ref{fig:graph_construction_sourcemaps}a).
While the state dependent switching fields
$U_i^\pm(S)$ define for each state and value of $U$ whether its hysterons are stable, they do not specify what happens when a hysteron becomes unstable. Here we first consider the stability range of states $S$ and identify the critical hysterons that lose stability when $S$ becomes unstable (sec.~\ref{ssec:sourcemap}). We then
discuss the ensuing state transitions, which can take on the form of multi-step avalanches (sec.~\ref{ssec:transitions}), and {finally} discuss the possibility that such transitions are ill-defined (sec.~\ref{ssec:ill}).
\subsubsection{Stability, Scaffold and Scrambling}\label{ssec:sourcemap}
\begin{figure}[t]
	\centering
	\includegraphics[width=.9\textwidth]{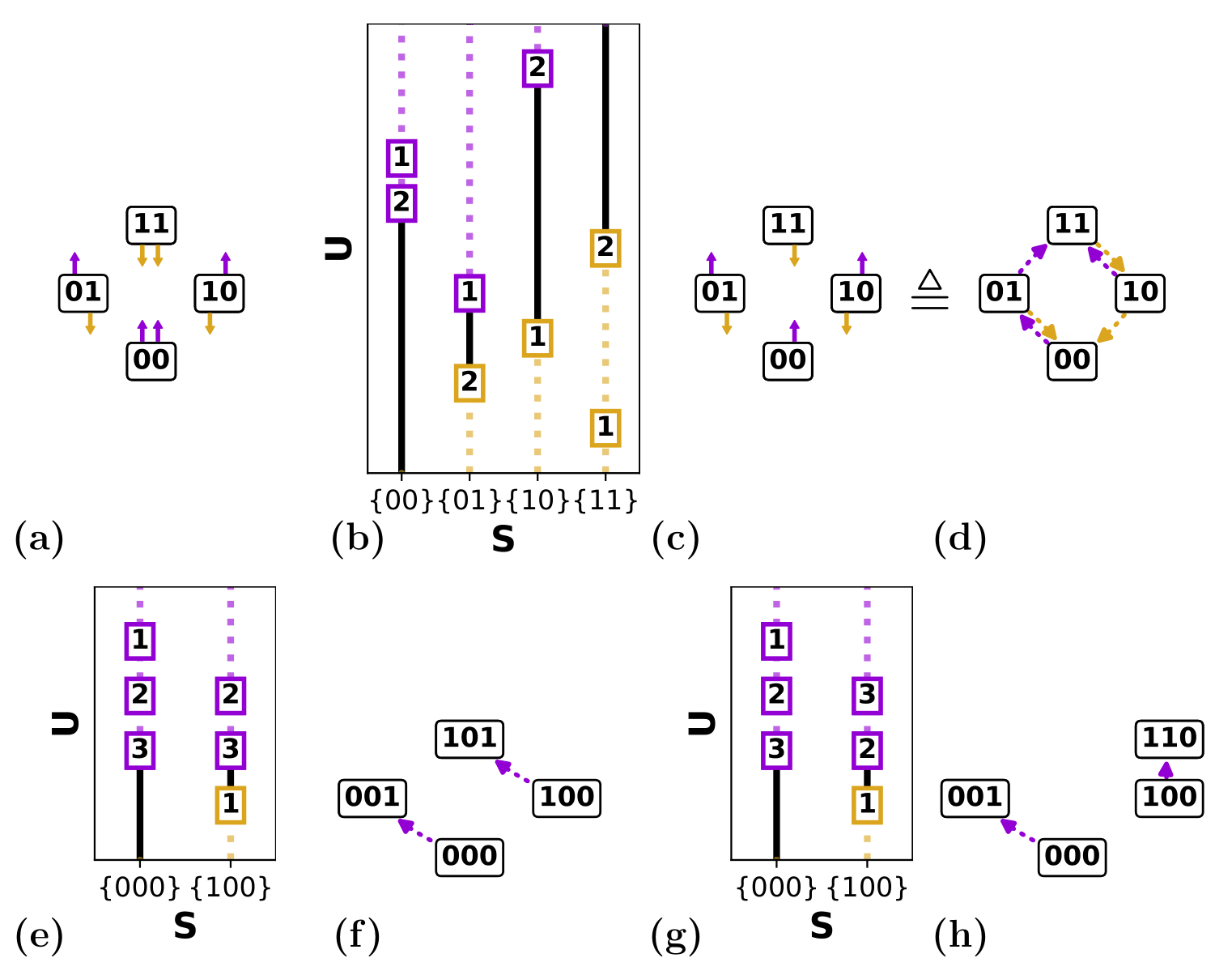}
	\caption{Scaffolds and switching fields. (a) Graphical representation of the set of $n=2$ switching fields: hysterons in state $s_i=0$ ($s_i=1$) have an up (down) switching field
		indicated by purple (gold) arrows. (b) Example of a given set of switching fields (boxes; number indicates hysteron index) and corresponding stability ranges (black lines) -- in the dashed regions, the state is unstable. (c) Corresponding scaffold, with the $2^{n+1}-2$ critical hysterons $k^\pm(S)$ indicated. (d) Alternative yet equivalent graphical representation of the scaffold showing {'passages', i.e.,} tentative transitions to a landing state $S^{(1)}$ that would occur upon flipping each critical hysterons $k^\pm(S)$ (see text).
		(e) Example of a subset of $n=3$ switching fields for the Preisach model where
		$\Delta_i^\pm(S)=0$. Since  $U_2^+(000)=U_2^+(001)=u_2^+$, and $U_3^+(000)=U_3^+(001)=u_3^+$,
		hysterons 2 and 3 flip in the same order for both states, and  since $k^+(000) = 3$, $k^+(100)=3$. (f) Graphical representation of the corresponding scaffold. (g) Example of a subset of $n=3$ switching fields, for the states $(000)$ and $(100)$, for the general model where $\Delta_i^\pm(S) \ne 0$. Since the switching fields $\{U_i^\pm(S)\}$ are independent, the order in which hysterons 2 and 3 flip may be different ('scrambled') between states $(000)$ and $(100)$. (h) Graphical representation of the corresponding scrambled scaffold.}
	\label{fig:graph_construction_sourcemaps}
\end{figure}
As a first step in defining the response of a collection of hysterons to variations of the global driving field $U$, we define here
the stability range of a state and its critical hysterons. We then introduce the concept of  the {\em scaffold}, and use it
to precisely define the concept of scrambling \cite{vhecke2021,bense2021,szulcmunganregev2022cooperative}.

\begin{definition}[{\em Stability range and state switching fields}]\label{def:state_sfs}
	Let $S$ be the state of a collection of hysterons with state-dependent switching fields $\{U_i^\pm(S)\}$. State $S$ is stable if $U$ is smaller than all
	its up switching fields $U_i^+(S)$ and larger than all its down switching fields
	$U_i^-(S)$. Hence, $S$ has a {\em stability range}
	$[U^-(S), U^+(S)]$, where $U_i^\pm(S)$ are the {\em state switching fields} (Fig.~\ref{fig:graph_construction_sourcemaps}b):
	\cite{mungan2019structure}:
	\begin{equation}\label{eq:global}
		\begin{split}
			U^+(S) &= \min U_{i}^+(S)~,\\
			U^-(S) &= \max U_{i}^-(S) ~.\\
		\end{split}
	\end{equation}
\end{definition}
When $S$ is initially stable and $U$ is swept up (down) to $U^+(S)$
($U^-(S)$), $S$ loses stability through the instability of
a single hysteron. We use the following definition for these hysterons:

\begin{definition}[{\em Critical hysterons and scaffold}]\label{def:critical}
	Let $S$ be the state of a collection of hysterons with state-dependent switching fields $\{U_i^\pm(S)\}$. The {\em critical hysteron} $k^+(S)$ ($k^-(S)$) is the hysteron which has the lowest up (highest down) switching field at state $S$:
	\begin{equation}\label{eq:global_indices}
		\begin{split}
			k^+(S) &= \arg\min U_{i}^+(S) ~,\\
			k^-(S) &= \arg\max U_{i}^-(S)~.\\
		\end{split}
	\end{equation}
	We refer to the collection of critical hysterons $\{k^\pm(S)\}$ at all states for a given set of switching fields as the {\em scaffold} of the hysteron system.
\end{definition}

The saturated states  have only one critical hysteron, 
whereas
all other  states have two critical hysterons. 
We 
note that persistently unstable states may arise when $U^+(S)<U^-(S)$; such states
can play a role as intermediate states in avalanches 
(section~\ref{ssec:transitions}).

The definitions (\ref{def:state_sfs}-\ref{def:critical})
map the $n\cdot 2^n$ hysteron switching fields $\{U_i^\pm(S)\}$
to $2^n-2$ critical hysterons $\{k^\pm(S)\}$ and
corresponding state switching fields $\{U^\pm (S)\}$. We note that the scaffold defined in Def. \ref{def:critical} is an important object: as its name suggests, it provides the underlying structure on which state transitions will be defined (Fig.~\ref{fig:graph_construction_sourcemaps}c).
This key role 
motivates an alternative yet equivalent graphical representation of the scaffold, where
we show tentative transitions 
associated with 
flipping the critical hysterons.
We term these tentative transitions {\em passages}.
(Fig. ~\ref{fig:graph_construction_sourcemaps}d). We stress here that passages are {\em not} necessarily equal to state transitions, and we
discuss the relation between scaffold and transitions in detail in section {\ref{ssec:transitions}}.\\

We now employ the scaffold to clarify the recently introduced {\em scrambling} property \cite{vhecke2021, bense2021,szulcmunganregev2022cooperative}. Loosely speaking, scrambling
was introduced for pairs of transitions that 
evidence hysteron interactions \cite{vhecke2021},
but its definition becomes cumbersome
when avalanches are present. As we show below, the scaffold allows a precise definition that does not suffer from such subtleties.

We will give the formal definition of scrambling first, as this allows us to make a clear argument for why scrambling is not possible in the Preisach model. We then illustrate our argument with specific examples of a scaffold in the Preisach model and of a scrambled scaffold.
\begin{definition}[{\em Scrambling}]
	Let $\{U_i^\pm(S)\}$ be the set of switching fields of a collection of hysterons, and let $\{k^\pm(S)\}$ be the corresponding scaffold. Let $k^+(S_A), k^+(S_B)$ be the critical up hysterons for a pair of distinct states $S_A$, $S_B$. The pair of critical hysterons $k^+(S_A), k^+(S_B)$ is {\em scrambled} if:
	\begin{equation}
		\begin{aligned}
			k^+(S_A) \in I_0(S_B) ~, \\
			k^+(S_B)\in I_0(S_A) ~, \\
			k^+(S_A)  \neq k^+(S_B) ~. \\
		\end{aligned}
	\end{equation}
	Similarly, a pair of critical down hysterons $k^-(S_A), k^-(S_B)$ is scrambled if:
	\begin{equation}
		\begin{aligned}
			k^-(S_A) \in I_1(S_B) ~, \\
			k^-(S_B)\in I_1(S_A) ~, \\
			k^-(S_A)  \neq k^-(S_B) ~.\\
		\end{aligned}
	\end{equation}
\end{definition}

Crucially, we now state the following:

\begin{corollary}\label{theorem:scrambling}
	The scaffold $\{k^\pm(S)\}$ for a set of switching fields of the form $U_i^\pm(S) = u_i^\pm$ cannot have scrambling.
\end{corollary}

To see that Corollary  \ref{theorem:scrambling} is true, consider a pair of hysterons $i$ and $j$. If $k^+(S_A) = i$, and $s_j = 0$, then by Definition \ref{def:critical}, $U_i^+(S) < U_j^+(S)$. Thus, in the Preisach model, $u_i^+ < u_j^+$. Yet, if there is another state $S_B$ where $k^+(S_B)$ and $s_i = 0$, then by the same reasoning, $u_i^+ > u_j^+$. Therefore, this situation cannot occur in the Preisach model.

A practical interpretation of scrambling is that in the Preisach model, the critical hysterons for different states are tightly connected via the order of the individual hysteron switching fields $u_i^\pm$. For general state-dependent switching fields $U_i^\pm(S)$, there is no such constraint. We illustrate this in the example below.

\begin{example}
	Suppose that we have a collection of non-interacting hysterons which has $k^+(000) = 2$. In the Preisach model, this implies  that $u_2^+ < u_3^+$, and therefore imposes that $k^+(100) = 2$ (Fig.~\ref{fig:graph_construction_sourcemaps}e-f). In contrast, now suppose that the switching fields $U_i^\pm(S)$ can be chosen entirely independently. In this case, the fact that $k^+(000)=2$ requires that $U_2^+(000) < U_3^+(000)$ does not impose any constraints on $U_2^+(100), U_3^+(100)$, so that it is possible to have $k^+(100)=3$ in the same scaffold (Fig.~\ref{fig:graph_construction_sourcemaps}g-h).
\end{example}

In summary, scrambling is an effect of hysteron interactions which leads to scaffolds that are not possible in the Preisach model. We note that scrambling can only occur for $n \geq 3$, and define a scaffold as scrambled when it
contains at least one pair of scrambled critical hysterons.

\subsubsection{Transitions and Avalanches}\label{ssec:transitions}
\begin{figure}[t]
	\includegraphics[width=.9\textwidth]{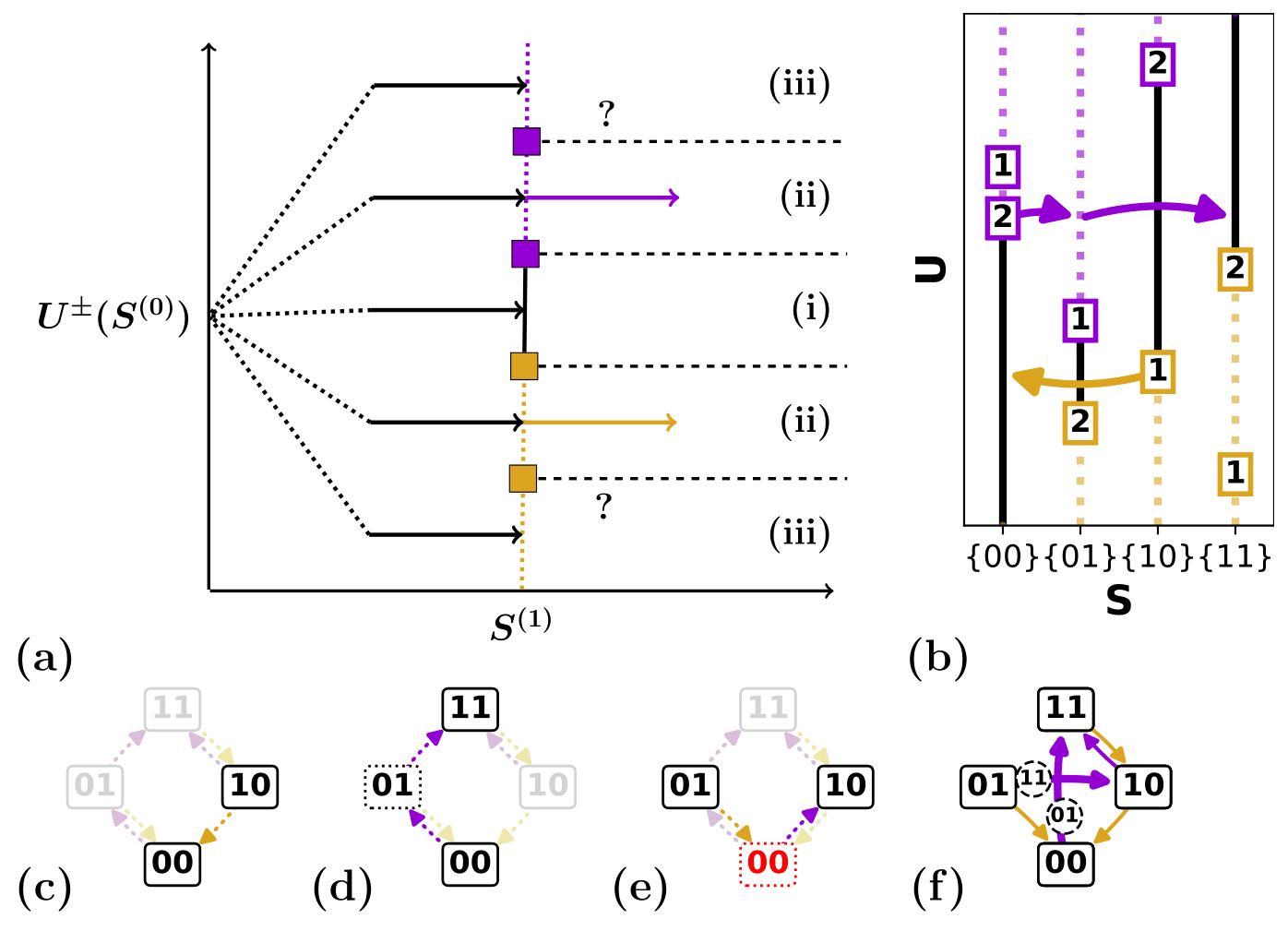}
	\caption{
		Construction of transitions and full transition graph from the set of switching fields. (a) Schematic of the three scenarios for the landing state $S^{(1)}$ at driving field $U^\pm(S^{(0)})$: (i) $S^{(1)}$ is stable; (ii) a single hysteron in state $S^{(1)}$ is unstable; (iii) $S^{(1)}$ multiple hysterons in $S^{(1)}$ are unstable (see text). (b) Two
		transitions ($l=1$ transition $(10)\downarrow (00)$ and $l=2$ transition $(00)\uparrow(01)\uparrow(11)$) for the set of switching fields shown in
		Fig.~\ref{fig:graph_construction_transitions}b.
		(c) The $l=1$ transition $(10)\downarrow(00)$ follows the scaffold (faded) (d)
		The avalanche transition $(00)\uparrow(01)\uparrow(11)$ follows the scaffold.
		(e) The {tentative} transition $(01)\downarrow(00)\uparrow(10)$ is incompatible with the scaffold, as the transition requires that $k^+(00) = 1$, while the scaffold specifies that $k^+(00) = 2$.
		(f) Full t-graph for the set of switching fields shown in panel b.
	}
	\label{fig:graph_construction_transitions}
\end{figure}
When state $S$ becomes unstable, due to either an up sweep or a down sweep of $U$, this triggers a transition to a new stable state $S'$, which we denote as $S \rightsquigarrow S'$.
Here we discuss in detail how
the system evolves from $S$ to $S'$, how $S'$ is determined, and under which conditions the transition and $S'$ are well-defined.

Each transition is initiated by the flipping of one of the critical hysterons $k^\pm(S)$, leading to a new 'landing' state $S^{(1)}$. In the Preisach model, the landing state is always stable at $U_k^\pm(S)$ \cite{terzimungan2020,mungan2019structure, ferrari2022preisach}, so that $S'=S^{(1)}$.
This produces the transition $S \rightsquigarrow S^{(l)}$ which trivially follows from the passages in the scaffold - in fact, for the Preisach model, transitions are equal to passages, and the scaffold captures all transitions.

In the presence of hysteron interactions, the stability of the landing state $S^{(1)}$ is no longer guaranteed, and this may lead to multi-step {\em avalanches}, which proceed via a sequence of intermediate states. We denote
such transitions as
$S^{(0)} \rightarrow  S^{(1)} \rightarrow \dots \rightarrow S^{(l)}$, where $S^{(0)}$ is the initial state,
$S^{(1)}$ is the landing state,
$S^{(1)}-S^{(l-1)}$ are intermediate states
and $S^{(l)}$ is the final state, and define the transition length as $l$.
Note that, in our notation, straight arrows represent flips of single hysterons, whereas the squiggly arrows represent a full transition 
$S^{(0)} \rightsquigarrow S^{(l)}$.

We now show how a single transition is constructed, given the set of switching fields. A transition is initiated when $k$ flips at $U=U_k^\pm(S)$, yielding the first step $S^{(0)}\rightarrow S^{(1)}$. There are three possible scenarios (Fig.~\ref{fig:graph_construction_transitions}a) depending on the stability of the landing state $S^{(1)}$:
\begin{enumerate}
	\item $S^{(1)}$ is stable at $U_k^\pm(S^{(0)})$;
	\item A single hysteron in state $S^{(1)}$ is unstable at $U_k^\pm(S)$;
	\item Multiple hysterons in state $S^{(1)}$ are unstable at $U_k^\pm(S)$.
\end{enumerate}
When the landing state $S^{(1)}$ is stable (case ({\em i})),
we obtain the $l=1$ transition $S^{(0)}	\rightsquigarrow S^{(1)}$.
For example, the switching fields shown in Fig.~\ref{fig:graph_construction_transitions}b produce the $l=1$ transition
$(10)\downarrow (00)$ (Fig.\ref{fig:graph_construction_transitions}b-c).

We now turn to case ({\em ii}), where a single hysteron $\kappa$ in state $S^{(1)}$ is unstable at $U=U_k^\pm(S)$. This provokes the
next step $S^{(1)}\rightarrow S^{(2)}$ {- please note that this scenario can occur even for a persistently unstable state (section \ref{sec:general})}.
For the state $S^{(2)}$, the same three scenarios can occur as for $S^{(1)}$:
if $S^{(2)}$ is stable (case ({\em i})),
the transition terminates. This produces the $l=2$ avalanche
$S^{(0)}	\rightsquigarrow S^{(2)}$, which proceeds
as  $S^{(0)}\rightarrow S^{(1)}\rightarrow S^{(2)}$.
We illustrate an example of an $l=2$ avalanche $(00)\uparrow(01)\uparrow(11)$ in Fig.~(\ref{fig:graph_construction_transitions}b,d).
If one of the hysterons in $S^{(2)}$ is unstable (case ({\em ii})), the transition proceeds to the next state $S^{(3)}$. Assuming that case ({\em iii}) does not occur and that states are not revisited
- as discussed below, both scenarios lead to ill-defined transitions\cite{vhecke2021} -  we see that avalanches are constructed iteratively.

We now show that the scaffold (Definition ~\ref{def:critical}) plays an important role in determining possible avalanches.

\begin{theorem}[{\em Relation between scaffold and avalanches}]\label{theorem:scaffold}
	Let $S^{(0)} \rightarrow S^{(1)} \rightarrow \dots \rightarrow S^{(l)}$ be a transition consistent with a set of switching fields $\{U_i^\pm(S)\}$, which has the corresponding scaffold $\{k^\pm(S)\}$. If the transition proceeds via a sequence of single hysteron flips $\kappa_0, \kappa_1, \dots ,\kappa_{l-1}$, each hysteron $\kappa_\lambda$ must be one of the critical hysterons $k^\pm(S^{(\lambda)})$.
\end{theorem}

It is easily shown that Theorem \ref{theorem:scaffold} is true via proof by contradiction. Suppose for definiteness that $\kappa_\lambda$ switches up, and $\kappa_\lambda \neq k^+(S^\lambda)$. Then, from Definition \ref{def:critical}, the switching field of hysteron $\kappa_\lambda$ is higher than that of $k^+(S^{(\lambda)})$. Thus, if $\kappa_\lambda$ is unstable, then $k^+(S^{(\lambda)})$ is unstable as well, which contradicts with the prerequisite that only a single hysteron flips at a time.

In essence, Theorem \ref{theorem:scaffold} states that avalanches must follow the passages of the scaffold. Hence, once the scaffold
is constructed from the switching fields, it immediately restricts the transitions
that can occur; for example,
for the scaffold shown in Fig.~\ref{fig:graph_construction_transitions}e,
the transition $(01)\downarrow(00)\uparrow(10)$ is forbidden. This relation between scaffold and avalanches can be explored to label an avalanche solely by its starting state and its sequence of up/down flips, letting the scaffold dictate the full transition path.
For example, the transition path $(00)\uparrow(01)\uparrow(11)$
shown in Fig. ~\ref{fig:graph_construction_transitions}d
is labeled as 00uu.
{Consequently}, all possible (avalanche) transitions can simply be collected in a set of binary trees;
we further elaborate on and make use of this scaffold/avalanche relation in sec.~\ref{sec:allgraphs}.

\subsubsection{Ill-defined transitions}\label{ssec:ill}
\begin{figure}[t]
	\includegraphics[trim = 0 0 50 0, width=.9\textwidth, clip]{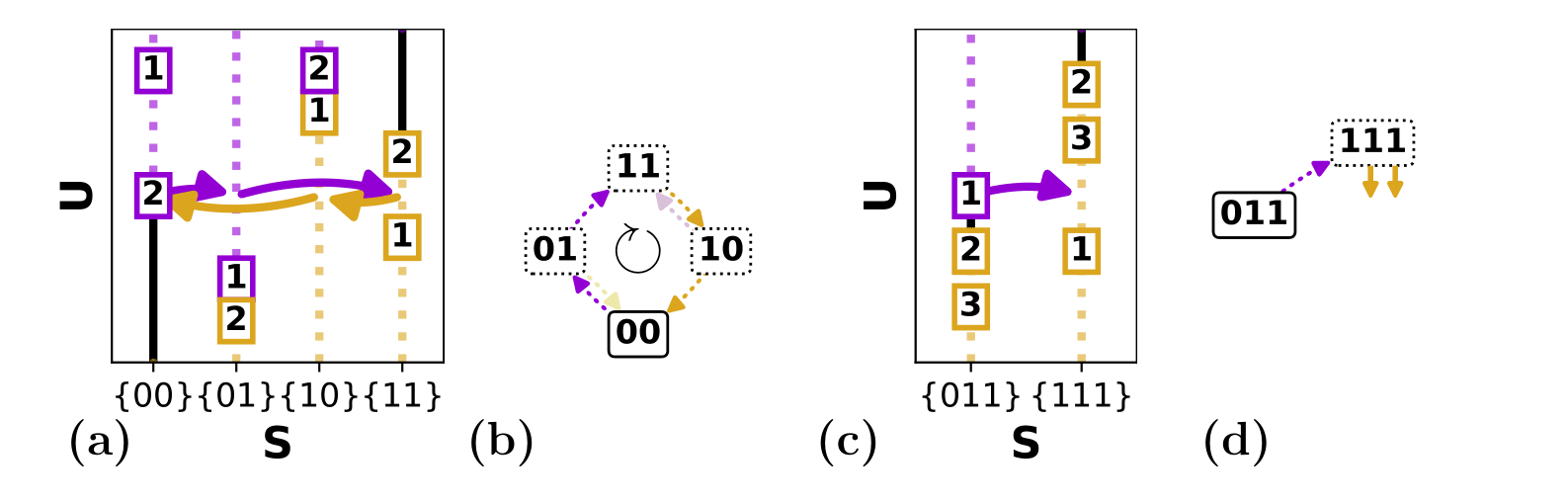}
	\caption{
		The two types of ill-defined transitions.
		(a) Set of switching fields that leads to a self-loop
		$(00)\uparrow (01)\uparrow (11)\downarrow (10) \downarrow (00)\uparrow \dots$.
		(b) Graphical representation for this self-loop.
		(c) Example of a (subset of) switching fields that leads to a race condition, due to the instability of multiple hysterons in state $(111)$ at $U^+(011)$.
		(d) Graphical representation for this race condition: we cannot draw a tentative down transition because the order in which hysterons 2 and 3 flip is not well defined.
	}
	\label{fig:ill_defined}
\end{figure}
So far, we have discussed how a given set of switching fields produces transitions. However, there are two mechanisms by which a set of switching fields
produce transitions that are ill-defined.

{First}, certain choices of switching fields produce self-loops, where after a number of steps, an avalanche revisits an earlier state \cite{baconnier2024proliferation}.In the simplest case, the switching hysteron is unstable in its landing state: for example, if after the partial transition $(00)\uparrow(01)$ hysteron 2 is unstable in $(01)$, this sets up a loop $(00)\uparrow(01)\downarrow(00)\uparrow(01)\downarrow\dots$. More generally, self-loops can arise from any cyclical path within the scaffold - for an example of a longer loop, see Fig. \ref{fig:ill_defined}a-b.
Such an orbit can never reach a stable state, and
such loops cannot occur in dissipative systems.
We consider the hysteron model ill-defined for switching fields that produce such self-loops\cite{vhecke2021}.

{Second}, when a transition reaches an intermediate state $S^{(\lambda)}$ where more than one hysteron is unstable at the critical field $U^\pm(S^{(0)})$, we consider the transition ill-defined - for an example of such case ({\em iii}) scenario,
see Fig.~\ref{fig:ill_defined}c-d. The problem is that
when multiple hysterons are unstable, the sequence of hysteron flips becomes ill-defined.
This is because  flipping operations do not commute for interacting hysterons:
for example, when the hysterons $\kappa_1$ and $\kappa_2$ are unstable, and hysteron $\kappa_1$ flips first, this may make hysteron $\kappa_2$ stable again, whereas when hysteron $\kappa_2$ flips first, hysteron $k_1$ may remain unstable. Hence, different choices for the sequence of hysteron flips may then lead to different transition paths, a situation known as a critical race condition \cite{huffman1954}. 

We note that some authors resolve race conditions by simply demanding that each intermediate step in a transition only flips one hysteron, and picking the most unstable hysteron if there is more than one; for such a model, the scaffold/avalanche relation specified by Theorem ~\ref{theorem:scaffold} is maintained \cite{lindeman2021, keimpaulsen2021}.

Both self-loops and {race conditions} arise because models for interacting hysterons
have a very simple update rule.
More physically complete models, based on an energy landscape and physical dynamics, would not feature race conditions or loops -- transitions would be well defined and loops would be avoided due to dissipation. All in all, hysteron models are perhaps the starting point, but not the end point for studies of the sequential response of complex media. Nevertheless, studies of more complex models come at a significant computational expense, and despite their (over)simplicity, hysteron models have proven valuable in capturing experimental and numerical data, as well as giving insight in memory effects.

\subsection{Transition Graphs}\label{ssec:transitiongraphs}
All ingredients are now in place to construct the transition graph (t-graph) which encodes
the full driving response for a set of hysteron switching fields $U_i^\pm(S)$.
To do so, we first construct the scaffold (section \ref{ssec:sourcemap}), and then
iteratively construct the full transition path $S^{(0)}\rightarrow S^{(1)}\rightarrow\dots\rightarrow S^{(l)}$
for each up and down transition. We collect these transitions in a directed graph, where the states form the nodes, and the transitions $S^{(0)} \rightsquigarrow S^{(l)}$ form the edges. {Furthermore,} the intermediate states $S^{(1)}, \dots, S^{(l-1)}$ {are saved as edge attributes. 
	
	As an example,
	the set of switching fields $\{U_i^\pm(S)\}$ shown in Fig.~\ref{fig:graph_construction_transitions}b produces the well-defined t-graph shown in Fig.~\ref{fig:graph_construction_transitions}f.
	In summary, the driving response of a collection of $n$ hysterons, characterized by a set of switching fields $\{U_i^\pm\}$, is captured by a directed t-graph.
	
	We note that our t-graphs contain more information than what is usually considered.
	First, for a given set of switching fields
	the intermediate states are fully specified, whereas in, e.g., experimental contexts intermediate states in an avalanche can typically not be observed \cite{bense2021}. Second,
	our t-graphs may contain so-called  Garden-of-Eden (GoE) states that are not reachable from the saturated states (for an example see state $(01)$ in Fig.~\ref{fig:graph_construction_transitions}f) \cite{vhecke2021, juleslechenault2021}.
	Unless otherwise specified, we will deal with t-graphs where GoE states and intermediate states are both included. In our visualization of t-graphs, we thus not only  indicate a transition's direction (up/down) and length $l$\cite{vhecke2021} but also explicitly indicate the intermediate states.
	
	\section{Graph Design}\label{sec:graphdesign}
	In this section we consider the design question, also referred to as the inverse problem: given a transition graph, or part of a transition graph, what are the necessary and sufficient conditions on the switching fields so that they
	lead to this (sub)graph? Consistent with earlier work on this inverse problem
	\cite{vhecke2021, keimpaulsen2021},
	we find that the design conditions take the form of sets of inequalities
	of the switching fields. Here, we establish a systematic approach for constructing and utilizing these design inequalities. We first present their systematic derivation (section \ref{ssec:design_ineqs}), then discuss how the design inequalities specify a partial order on the switching fields (section \ref{ssec:porder}). In particular, we utilize the inequalities
	to determine if a given t-graph topology is realizable (section \ref{ssec:existence}), and finally discuss how to construct explicit sets of switching fields that realize specific (sub)-graphs (section \ref{sec:solve}).
	\subsection{Design Inequalities}\label{ssec:design_ineqs}
	\begin{figure}[t]
		\large
		\includegraphics[width=.9\textwidth]{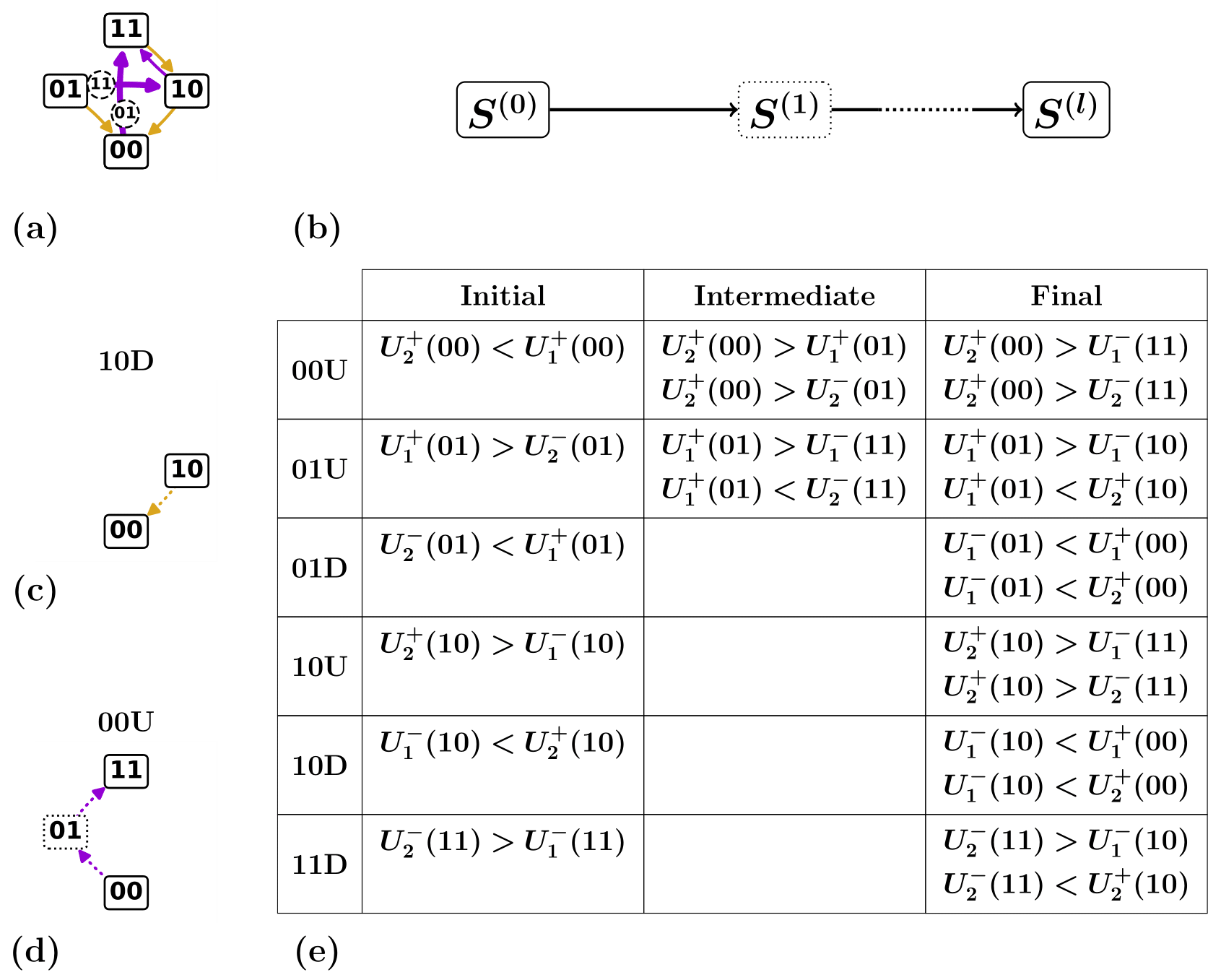}
		\caption{Design inequalities.
			(a) Example of an $n=2$ target t-graph.
			(b) The design inequalities are organized in three groups
			pertaining to the initial state $S^{(0)}$, intermediate states $S^{(1)}, \dots, S^{(l-1)}$, and final state $S^{(l)}$.
			(c) Length-1 transition $(10)\downarrow (00)$. (d) Length-2
			transition $(00)\uparrow (01)\uparrow(11)$. (e) The design inequalities for the target graph shown in panel (a). We label transitions by
			their starting state $S^{(0)}$ and their up or down direction:
			for example, the up transition starting from state $(00)$ is labeled 00U, and the down transition starting from state $(10)$ is denoted
			10D (see also panels (c) and (d)). Note that all inequalities are organized by transition, and can be {further} separated {by} whether they are given by the initial state (scaffold), intermediate state (if present), or final state inequalities.
		}
		\label{fig:design_ineqs}
	\end{figure}
	
	We define the {\em design inequalities} as the necessary and sufficient conditions on the switching fields $\{U_i^\pm(S)\}$ that produce
	a specific (sub)graph.
	We will frequently illustrate our approach with specific examples, such as the t-graph
	and its scaffold shown in Fig.~\ref{fig:design_ineqs}{a}.
	As we show, each transition in a t-graph corresponds to a set of design inequalities that result from {\em (i)}
	conditions on the initial state (its stability and critical hysterons),
	{\em (ii)}
	the stability of the final state, and, if the transition is an avalanche,
	{\em (iii)}
	the (in)stabilities and critical hysterons of the intermediate states.
	Combining the {individual} design inequalities for multiple transitions, taking potential redundancies into account, produces {the set of inequalities} for {a} specific
	{t-graph} or {part} thereof.
	
	\subsubsection{Conditions for Single Transitions}
	We now formulate the design inequalities for a single transition.
	As avalanches for which only the initial state $S^{(0)}$ and final state $S^ {(l))}$ are specified may proceed along various paths of intermediate states, each producing a different set of design inequalities, we assume
	that the full transition path $S^{(0)}\rightarrow S^{(1)} \rightarrow \dots \rightarrow S^{(l))}$ is known (Fig.~\ref{fig:design_ineqs}{b}). {We illustrate that} these inequalities readily emerge by constructing transitions as in section \ref{sec:drive}, and keeping track of the requirements that the switching fields must obey at each step.
	
	\paragraph{Initial inequalities.--}
	The first set of design inequalities follows from the required stability of
	state $S^{(0)}$ for some range of $U$, and from
	the critical hysteron $k$ that flips when
	the driving $U$ is increased or decreased.
	For hysteron $k$ in phase 0 (1) to be the critical hysteron that initiates {an up (down) transition $S^{(0)} \rightsquigarrow S^{(l)}$}, the set of switching fields $\{U_i^\pm(S)\}$ must obey the set of inequalities:
	\begin{align}
		\mathrm{up: }  &&  U_{k}^+(S^{(0)}) &< U_i^+(S^{(0)}) ~ \forall i\in I_0(S^{(0)})\backslash\{k\}~,\\
		\mathrm{down: } &&  U_{k}^-(S^{(0)}) &> U_i^-(S^{(0)}) ~ \forall i\in I_1(S^{(0)})\backslash\{k\}~,
	\end{align}
	where $ I_{0/1}(S^{(0)})\backslash\{k\}$ is the collection of hysterons in phase $0~(1)$, unequal to $k$ in state $S^{(0)}$.
	
	Moreover, for $S^{(0)}$ to be initially stable in the case of an up (down) transition, all down (up) switching fields must be below (above) the critical switching field $U_k^\pm(S^{(0)})$:
	\begin{align}
		\mathrm{up: } && U_{k}^+(S^{(0)}) &> U_i^-(S^{(0)}) ~  \forall i\in I_1(S^{(0)})~,\\
		\mathrm{down: } && U_{k}^-(S^{(0)}) &< U_i^+(S^{(0)}) ~ \forall i\in I_0(S^{(0)})~.
	\end{align}
	
	\begin{example}
		In
		Fig.~\ref{fig:design_ineqs}{a}, the {condition for the up transition from state $(00)$ to be initiated by critical hysteron $k^+(00)=2$} is $U_2^+(00)<U_1^+(00)$, and the stability condition is trivially satisfied. Similarly, {the down transition from state $(11)$} produces one critical hysteron condition and no stability conditions. In contrast, the up and down transitions from the
		states $(01)$ and $(10)$ specify one stability condition and no conditions for the critical hysterons.
	\end{example}
	
	\paragraph{Final inequalities.--}
	A second set of inequalities arises from the requirement that the final state $S^{(l)}$ is stable. For the final state $S^{(l)}$ {of a transition $S^{(0)} \rightsquigarrow S^{(l)}$} to be stable at the critical driving ${U_0:=}U_k^+(S^{(0)})$ where it is initiated, a set of switching fields $\{U_i^\pm(S)\}$ must obey the set of inequalities:
	\begin{align}
		{U_0} &< U_i^+(S^{(l)}) ~\forall i\in I_0(S^{(l)})  ~,\\
		{U_0} &>U_i^-(S^{(l)}) ~ \forall i\in I_1(S^{(l)}) ~,
	\end{align}
	{For a down avalanche at $U{_0}=U_k^-(S{^{(0)}})$, the equations are identical.}

	\begin{example}
		For the transition $(10)\downarrow(00)$
		(Fig.~\ref{fig:design_ineqs}{c}), the stability of the final state $(00)$ at $U_1^-(10)$ yields:
		\begin{align*}
			U_1^-(10) &< U_1^+(00) ~,\\
			U_1^-(10) &< U_2^+(00) ~.
		\end{align*}
	\end{example}
	
	\paragraph{Intermediate inequalities.--}
	While $l=1$ transitions, like {$(10)\downarrow(00)$} in the example of Fig.~\ref{fig:design_ineqs},
	only produce initial state and final state inequalities,
	each intermediate step in an avalanche of length $l>1$
	produces additional conditions to ensure that each
	intermediate step is well-defined and proceeds as described.
	Consider
	{again} an avalanche initiated at {$U{_0}=U_k^+(S{^{(0)}})$ (for an up transition) or $U{_0}=U_k^+(S{^{(0)}})$ (for a down transition). We now look at at} an intermediate state $S^{(\lambda)}$ where for definiteness hysteron $\kappa$ switches up, so $\kappa = k^+(S^{(\lambda)})$ For
	hysteron $\kappa$ to be unstable at $U_k^\pm(S)$, while all other hysterons are stable, a set of switching fields $\{U_i^\pm(S)\}$ must obey the set of inequalities:
	\begin{align}
		{U_0} &< U^+_i(S^{(\lambda)}) ~ \forall i\in I_0(S^{(\lambda)})\backslash\{\kappa\}~,\\
		{U_0} &> U_i^-(S^{(\lambda)}) ~ \forall i\in I_1(S^{(\lambda)})~,\\
		{U_0} &> U_\kappa^+(S^{(\lambda)}) ~.  \label{eq:criticalup}
	\end{align}
	Analogously, when hysteron $\kappa$ switches down, $\{U_i^\pm(S)\}$ must obey the inequalities:
	\begin{align}
		{U_0} &< U^+_i(S^{(\lambda)}) ~ \forall i\in I_0(S^{(\lambda)})~,\\
		{U_0} &> U_i^-(S^{(\lambda)}) ~ \forall i\in I_1(S^{(\lambda)})\backslash\{\kappa\}~,\\
		{U_0} &< U_\kappa^-(S^{(\lambda)}) ~.  \label{eq:criticaldown}
	\end{align}

	\begin{example}
		Consider the $l=2$ avalanche  {$(00)\uparrow(01)\uparrow(11)$}(Fig.~\ref{fig:design_ineqs}d).
		The instability of hysteron 1 at the intermediate state $(01)$ gives rise to the equations:
		\begin{align*}
			U_2^+(00) &> U_1^+(01)~,\\
			U_2^+(00) &> U_2^-(01)~.
		\end{align*}
	\end{example}
	
	Note that between the cases where $\kappa$ switches up and down, the inequalities involving the non-critical hysterons stay the same, and the only difference is a sign change in the inequality that involves $\kappa$ -- compare \eqref{eq:criticalup} and \eqref{eq:criticaldown}. 
	The number of inequalities of each type is fixed:  in general, there are $n-1$ initial state inequalities, which arise from the comparison of the critical hysteron $k$ against every other hysteron. {In addition there are $n$ final inequalities, which specify that each hysteron in the final state $S^{(l)}$ is stable.} {Similarly}, there are $n$ intermediate inequalities for each of the states $S^{(1)}, \dots, S^{(l-1)}$, which specify the stability of each hysteron in $S^{(\lambda)}$ under $U_k^\pm(S)$.
	
	\subsubsection{Full Graph}
	
	Combining the initial, intermediate and final inequalities, one obtains a set of inequalities that the switching fields must obey so that a given transition is realized. We see that, for example, the transition $(00)\uparrow (01)\uparrow(11)$ requires five inequalities: one initial inequality, two intermediate inequalities, and two final inequalities. In general, the number of inequalities per transition is $(l+1)n - 1$.
	
	This approach constructs conditions on the level of individual transitions, and is
	therefore modular. {To
		construct the necessary and sufficient conditions on the switching fields corresponding to a given t-graph or subgraph thereof, we simply
		combine the conditions of their respective transitions, as these graphs are nothing more than the collection of transitions from different initial states for a given $\{U_i^\pm(S)\}$.
		We note that there are often redundancies between these inequalities: for example, in Figure~\ref{fig:design_ineqs}e, the inequality $U_2^+(10) > U_2^-(11)$ appears twice.
		
		\subsection{Partial order and matrix representation}\label{ssec:porder}
		\begin{figure}[t]
			\includegraphics[width=.9\textwidth]{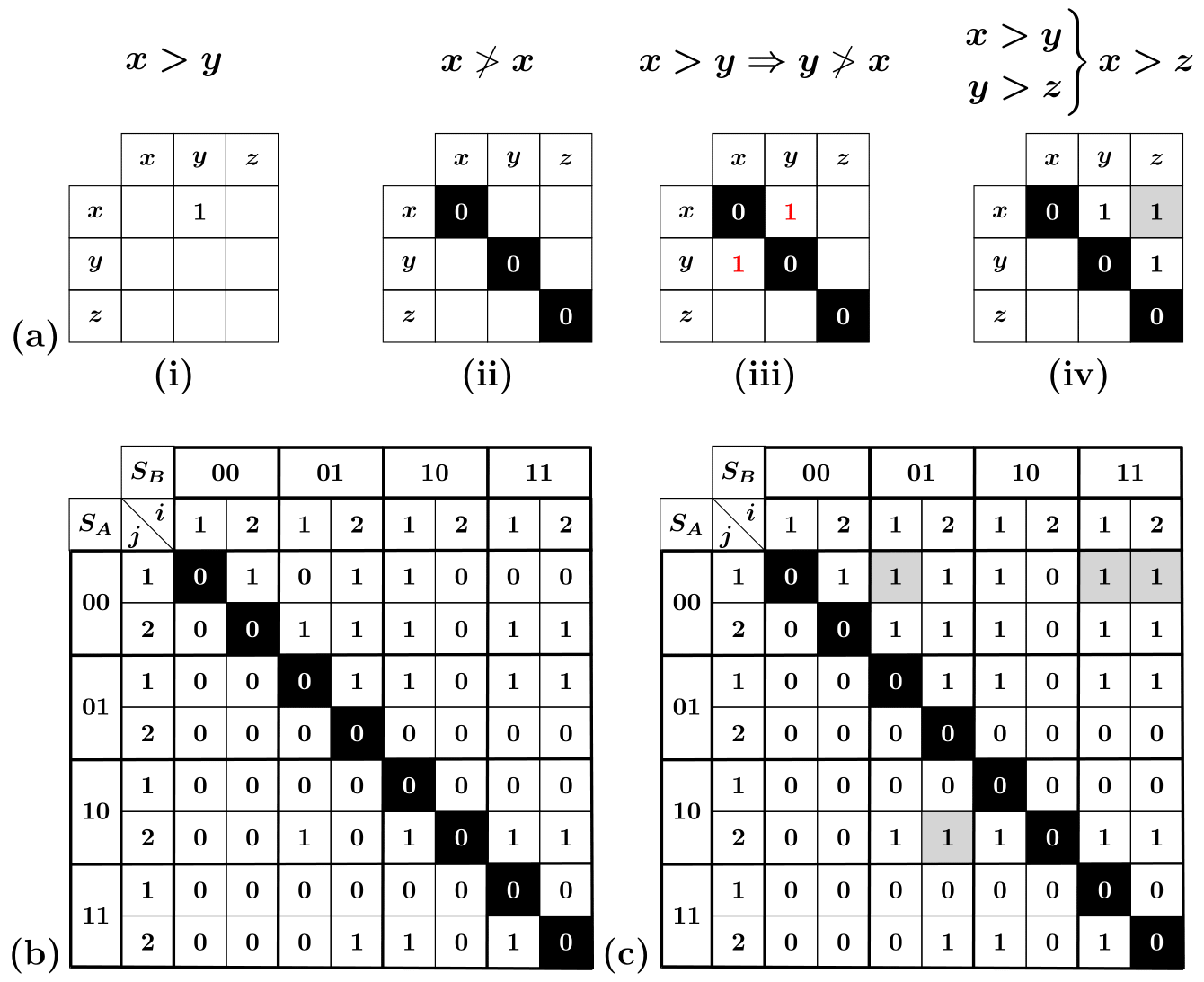}
			\caption{Partial order represented by adjacency matrices. (a) Representation of a partial order as an adjacency matrix (i) and three properties of a partial order: (ii) irreflexivity, shown by the diagonal highlighted in white-on-black; (iii) asymmetry, shown by a pair of entries highlighted in red, where $x>y$ rules out that $y>x$; and (iv) transitivity, where the induced inequality $x > z$ is highlighted in gray. (b) Example of {the list of design inequalities shown in Fig.~\ref{fig:design_ineqs}e in adjacency matrix form}. For the sake of readability, we index the switching fields $U_i^\pm(S)$ via the state $S$ and hysteron index $i$. The matrix thus contains $n\times n$ blocks of inequalities which concern the same states $S_A, S_B$ as shown by thicker lines. (c) Transitive closure of the adjacency matrix shown in (b); note the addition of induced inequalities, highlighted in gray.}
			\label{fig:adjmat}
		\end{figure}
		
		Considering the design inequalities from section~\ref{ssec:design_ineqs}, we observe 
		that they specify a {\em partial order} on the switching fields $U_i^\pm(S_j)$. To see this, note that the design inequalities are a collection of pairwise inequalities of the form $U_i^\pm(S_A) > U_j^\pm(S_B)$.
		{We briefly discuss how this helps our understanding of the design inequalities; for a more detailed overview of partial orders, we refer to established literature from discrete mathematics \cite{wallis2011beginner, grami2022discrete}.}
		
		We {choose to} represent a partial order as an adjacency matrix, where a 1 at position $(x, y)$ indicates the presence of a design inequality $x > y$, and a 0 indicates no relation (Fig. \ref{fig:adjmat}a-(i)).
		It is useful at this point to recall the general properties of partial orders -- irreflexivity, asymmetry, and transitivity. First,
		{\em irreflexivity} entails that $x \ngtr x$ (Fig. \ref{fig:adjmat}a-(ii)).  Second, a partial order relation is {\em asymmetric}, such that if $x > y$, then $y \ngtr x$ (Fig.~\ref{fig:adjmat}a-(iii)). Finally, a partial order has {\em transitivity}, meaning that if $x > y$ and $y > z$, then $x > z$ (Fig. \ref{fig:adjmat}a-(iv)).
		
		{It is straightforward to convert the inequalities tabulated in Fig.~\ref{fig:design_ineqs}f to a square matrix, where the rows and columns represent the $n \times 2^n$ switching fields (Fig. \ref{fig:adjmat}b). This conversion also immediately gets rid of redundancies - note that Fig.~\ref{fig:adjmat}b contains only seventeen inequalities whereas Fig.~\ref{fig:design_ineqs}f contained nineteen. This matrix is not yet sufficient to represent a partial order, however, because the third property -- transitivity -- is not satisified. To obtain a proper representation, we must construct a matrix that also includes all induced inequalities: the {\em transitive closure}\cite{grami2022discrete}. This is achieved by using a transitive closure algorithm such as Warshall's algorithm \cite{warshall1962}.}
		
		{We emphasize that, while the specific set of design inequalities that correspond to a given t-graph topology may depend on the algorithm used to generate it, the partial order associated with each t-graph is unique. Given that the design inequalities specify a {\em partial} order, though, they correspond to a discrete (and possibly very large) number of {\em total} orders of the switching fields. We further discuss this in section~\ref{sec:statweight}.}
		
		\subsection{Realizability of t-graphs}\label{ssec:existence}
		\begin{figure}[t]
			\centering
			\includegraphics[width=.9\textwidth]{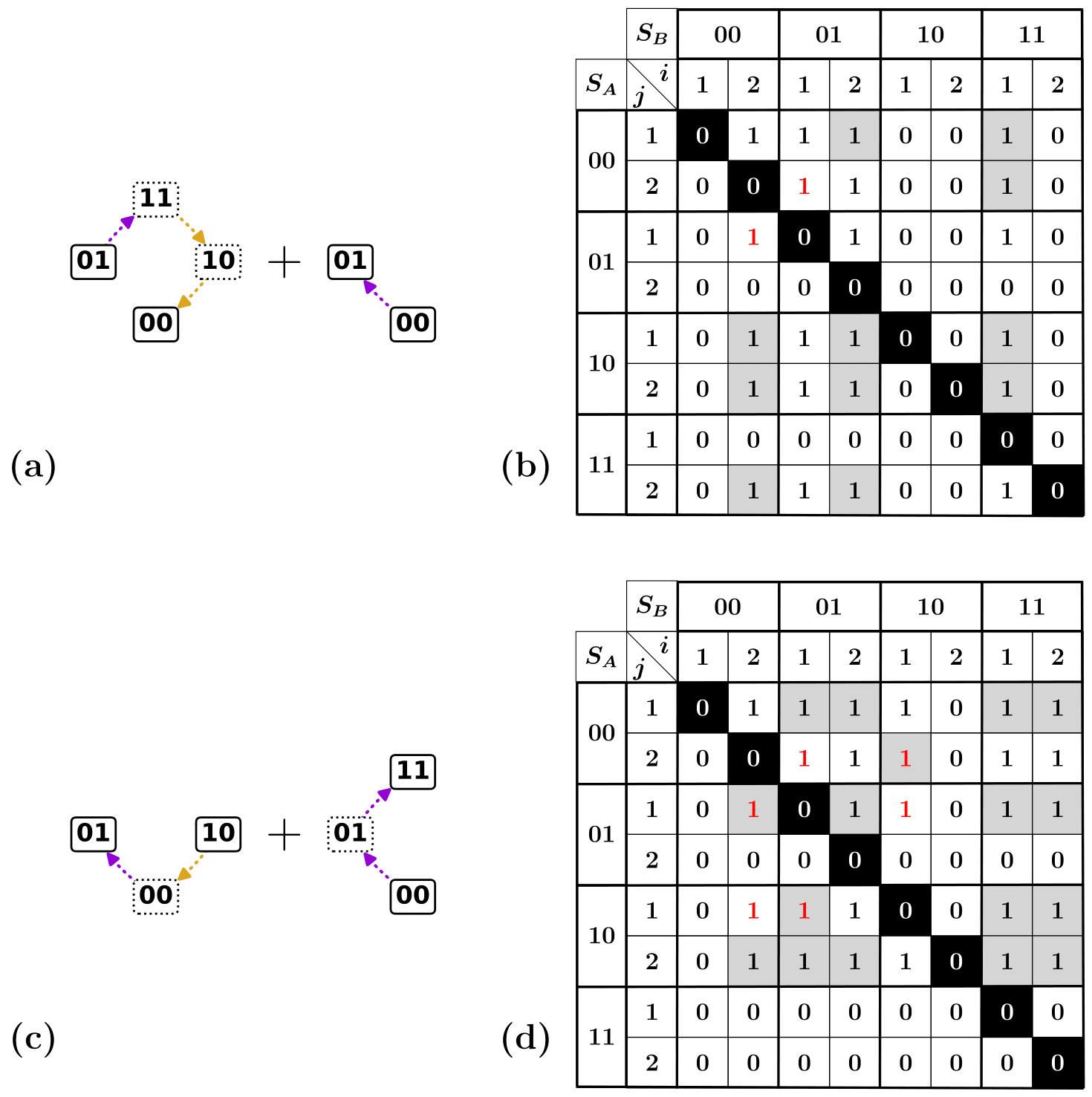}
			\caption{Illustration of impossible subgraphs and corresponding design inequalities. (a) Tentative combination of transitions $00\uparrow 01$ and $01\uparrow11\downarrow 10\downarrow 00$. (b) Corresponding design inequalities shown as an adjacency matrix, with entries induced via transitivity highlighted in gray. The entries (00, 2; 01, 1) and (01, 1; 00, 2)  show that asymmetry is broken (highlighted in red). (c) Tentative subgraph containing transitions $00\uparrow 01\uparrow 11$ and $10\downarrow 00\uparrow 01$. (d) Corresponding design inequalities. The three pairs of entries highlighted in red show that the asymmetry condition is broken; note that this breaking of asymmetry only becomes visible via the transitive closure (highlighted in gray).}
			\label{fig:impossible_subgraphs}
		\end{figure}
		
		While any target t-graph can be mapped to a set of design inequalities,
		{design inequalities may at times contradict each other}, such that they are not consistent with any set of switching fields.
		Hence, the question of existence of a set of switching fields that realize
		a target t-graph is tantamount to checking the satisfiability of the
		set of design inequalities. {To solve the general problem of checking whether a set of inequalities is consistent, one can use the classical method of Fourier-Motzkin elimination \cite{dantzig1972fourier}, or alternatively, more refined linear programming methods such as the simplex algorithm \cite{dantzig1997simplex}.}
		
		For our sets of pairwise inequalities, {checking solvability is more straightforward than in the general case:} inconsistencies {show up as} violations of asymmetry, i.e., pairs of inconsistent inequalities of the form $x>y$ and $x<y$ occurring simultaneously. We illustrate our approach by discussing two examples of non-realizable t-graphs.
		
		First we consider a target t-graph that contains a subgraph consisting of the $l=1$ up transition $(00)\uparrow (01)$, and the $l=3$ up transition $(01) \uparrow (11) \downarrow (10) \downarrow (00)$ (Fig.~\ref{fig:impossible_subgraphs}a).
		For this example, there is a direct contradiction between two of the design inequalities:
		the stability of the final state $(01)$ of the transition $(00)\uparrow (01)$ requires $U_2^+(00) < U_1^+(01)$, while the stability of the final state $00$ of the transition $(01) \uparrow (11) \downarrow (10) \downarrow (00)$ requires $U_2^+(00) > U_1^+(01)$ (Fig. \ref{fig:impossible_subgraphs}b). 
		
		Second, we consider Fig.~\ref{fig:impossible_subgraphs}c. For this example the contradiction
		in the design inequalities is not directly visible from the original design inequalities, but becomes manifest once the transitive closure is constructed.
		To see this, note that the up transition  $(00)\uparrow(01)
		\uparrow(11)$ requires that $U_2^+(00) > U_1^+(01)$, while the down transition $(01)\downarrow(00)\uparrow(10)$ requires $U_1^+(01) > U_2^-(10)$
		and  $U_1^-(10) > U_2^+(00) $. Together, this leads to the inconsistent chain of inequalities $U_2^+(00) > U_1^+(01)  > U_2^-(10) > U_2^+(00) $ (Fig. \ref{fig:impossible_subgraphs}d). Hence, this target graph can not be realized.
		
		{We emphasize that
			we focus here on realizability in a mathematical sense, i.e., asking if there is a  partial order of the switching fields that realizes a certain t-graph. The question of physical realizability is a separate issue. We note that for specific physical models, the switching fields can not be chosen arbitrarily. For example, when the switching fields arise from pairwise interactions \cite{vhecke2021,shohat2025geometric} this imposes constraints on the switching fields, and thus further limit the set of realizable t-graphs, that go beyond considerations of partial order. Other physical constraints that have been considered include interactions being symmetric~\cite{keimpaulsen2019,baconnier2024proliferation} and global coupling, which arises when e.g. hysterons are serially coupled~\cite{liu2024}. These physical considerations complicate the design inequalities beyond a mere partial order, such that one needs to fall back on generic linear programming techniques. However, the space of t-graphs as a whole can often be simplified for physical models, as we discuss in section~\ref{sec:allgraphs}. These models, while of great interest, deserve individual consideration, which goes beyond the general framework established here.}
		
		\subsection{Solving the design inequalities}\label{sec:solve}
		To conclude, we discuss how explicit solutions for the switching fields $\{U_i^\pm(S)\}$ can be constructed for a given {set of (consistent) design inequalities}. For our sets of pairwise inequalities, an explicit solution can be constructed by finding a
		total order (or 'linear extension' \cite{szulcmunganregev2022cooperative}) that satisfies the partial order specified by the design inequalities. We use a topological sorting algorithm such as Kahn's algorithm\cite{kahn1962topological, knuth1997art} to produce a random linear extension, without regard for the range and distributions of gaps between switching fields.
		
		To convert such a random linear extension to an explicit solution, we set the switching fields to be equidistant with a spacing of $\epsilon = {\big(n\cdot 2^n\big)}^{-1}$, and set the lowest switching field to $\epsilon/2$. This is a convenient choice because it guarantees that the switching fields lie within the range $[0, 1]$, and the system is in the saturated states $(00...)$, $(11...)$ at driving values of 0 and 1 respectively.
		
		Although we have emphasized that the design inequalities for a given t-graph topology only impose a partial order relation, additional constraints - for example, those imposed by the use of a specific model for $\Delta_i^\pm(S)$ - can lead to more complex sets of inequalities. In these cases, one can fall back on general linear programming methods {such as the simplex algorithm \cite{dantzig1997simplex}} to solve the full set of inequalities. {We note that, to make use of linear programming, one must first transform the set of strict design inequalities to a set of non-strict design inequalities $U_i^\pm(S_A) - U_j^\pm(S_B) \geq b$, with $b$ a small positive number. Like $\epsilon$ in the method described above, $b$ here sets an explicit spacing between switching fields, which is necessary because the problem is scale-invariant otherwise.}
		
		Altogether we have outlined   
		how for any t-graph or subgraph, we can construct corresponding design inequalities. These design inequalities allow us to check if a t-graph is realizable and, if so, to generate a random set of switching fields that realizes the graph. We make use of this in section \ref{sec:allgraphs} to generate all valid t-graphs for $n=2$. Moreover, the process of checking solvability and finding a solution is facilitated by the observation that the design inequalities only specify a partial order on the switching fields (section \ref{ssec:porder}). This observation has further implications for the statistical weight of t-graphs, as we will discuss in section \ref{sec:statweight}.}
	
	\section{Constructing and organizing all graphs}\label{sec:allgraphs}
	\begin{figure}[t]
		\includegraphics[width=.9\textwidth]{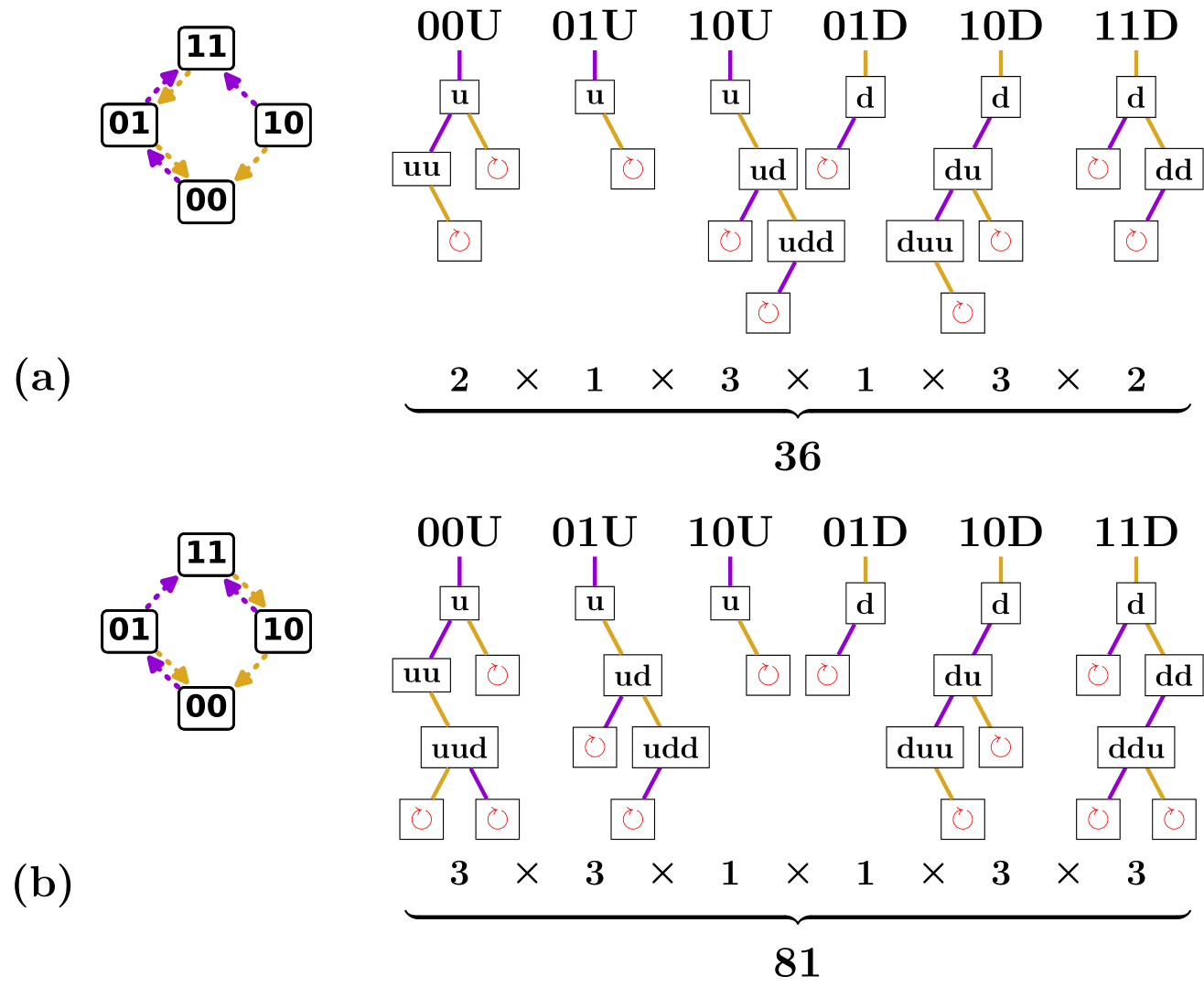}
		\caption{Binary trees of possible avalanches for $n=2$. a) First of two possible scaffolds for $n=2$, and associated binary trees of avalanches per transition. {We label transitions by
				their starting state $S^{(0)}$ and their up or down direction marked respectively as U or D (see also Fig.\ref{fig:design_ineqs})}. Each tree node represents a transition $S^{(0)}\rightarrow\dots\rightarrow S^{(l)}$, characterized by a starting state and up/down sequence{ - please note the difference between uppercase and lowercase labels, as U and D signify the initial direction of a transition, whereas u and d signify intermediate steps}. A length-$l$ transition can be extended to a transition of length $l+1$ by an up flip if $S^{(l)} =(00...)$, by a down flip if $S^{(l)}=(11...)$, and by either an up or down flip otherwise. Branches of the tree terminate where there is a self-loop, indicated in red. The number of avalanches in each tree is indicated below each transition; the product of these counts gives the number of avalanches for the given scaffold. b) Second of two possible scaffolds for $n=2$, and associated binary trees of avalanches per transition. Note that the structure of each tree is as in (a), but the nodes at which self-loops are encountered are different.}
		\label{fig:binary_trees}
	\end{figure}
	
	In this section we consider the space of all t-graphs.
	While the profusion of t-graphs with {$n\geq3$} makes brute force explorations
	unfeasible \cite{vhecke2021}, we build here on the observation that any
	t-graph is created by combining transitions
	which proceed on a scaffold (sections ~\ref{ssec:sourcemap},~\ref{ssec:transitiongraphs}). Thus, to systematically explore the space of all t-graphs, we proceed in three steps.
	First, for given $n$, we create and count all scaffolds (section~\ref{ssec:construct_scaffolds}).
	Second, for any state in a scaffold, all possible
	(avalanche) transitions can be organized by finite binary trees (section~\ref{ssec:binary_trees}).
	Third, by combining scaffolds and selecting transitions from the binary trees, all {\em candidate} t-graphs can be systematically generated. Only a fraction of these are consistent with the corresponding design inequalities, and these are the sought-after valid t-graphs.
	We use this  approach to determine all valid t-graphs for $n=2$, all scaffolds (and all avalanche-free t-graphs) for $n=3$,
	and a loose upper bound on the number of t-graphs for $n=3$ (section \ref{ssec:constructgraphs}).
	Our approach uncovers the root cause of the multiplicity and complexity of the space of all t-graphs: as each state allows a number of potential transitions, the multiplication of these numbers very quickly leads to astronomical numbers, while at the same time, {the combination of} longer and longer avalanches eventually leads to {many} t-graphs of which the design inequalities are inconsistent. {Thus, while the number of valid t-graphs for general $n$ is presumably much smaller than the number of candidate graphs, the complexity of the underlying design inequalities makes it challenging to systematically construct - or even enumerate - these graphs.}

	\subsection{Scaffolds}\label{ssec:construct_scaffolds}
	We now consider the multiplicity and organization of the scaffolds. To aid us in counting the scaffolds, we first define several useful attributes of the scaffold:
	
	\begin{definition}[{\em Magnetisation}]
		Let $S = (s_1, \dots, s_n)$ be a state of collection of $n$ hysterons. We define the
		{\em magnetization} $m:=\Sigma_i s_i$.
	\end{definition}
	\begin{definition}[{\em Up/down boundary and main loop}]
		For a given scaffold $k^\pm(S)$, the {\em up (down) boundary} is the sequence of up (down) passages that connect
		the saturated states $(0 0\dots )$ and $(1 1 \dots )$.
		
		The up and down boundary together form the {\em main loop} of the scaffold \cite{terzimungan2020, ferrari2022preisach}.
	\end{definition}
	
	We break {the} relabeling symmetry of the hysterons
	by requiring
	that the
	the up boundary of the main loop
	is fixed, so that the hysterons flip in the order $n, n-1, \dots, 1$ \cite{terzimungan2020}.
	With this convention, there are $n!$ possible down boundaries and hence $n!$ main loops,
	which can be labeled by the sequence of down transitions.
	For example, the main loop of the scaffold in Fig.~\ref{fig:binary_trees}a,
	$(00)\uparrow (01)\uparrow(11)\downarrow(01)\downarrow(00)$,
	can be denoted as $(1, 2)$ \cite{terzimungan2020, ferrari2022preisach}.
	
	The number $N_S(n)$ of scaffolds for $n$ hysterons can now be obtained from a simple combinatorial argument.
	For each up (down) transition, the number of possible critical hysterons $n^\pm(S)$ equals the number of hysterons in phase 0 (1), so that $n^+(S) = n-m$ and $n^-(S)=m$, where $m$ follows from $S$ straightforwardly.
	As the number of states with magnetisation $m$ equals $\binom{n}{m}$,
	we immediately find that
	\begin{equation}
		N_{S} = \frac{\prod_S (n^+(S)~ n^-(S))}{n!}=
		\frac{{\Big( \prod_{m=1}^n m^{\binom{n}{m}}\Big)}^2}{n!}~,
	\end{equation}
	where the division by $n!$ takes care of the relabeling symmetry.
	
	{The scaffolds
		can simply be labeled by the values of the critical hysterons, and can be organized by main loop and by the minimal number of scrambled passages. First, each main loop allows for the same number of scaffolds ($N_{S}/n!$) -- to see this, note that, irrespective of the main loop, the same amount of choices of up and down transitions at each value of $m$ are available. Second, for each main loop we can define one unscrambled or Preisach scaffold, where all critical transitions follow the order of the up and down transitions along the main loop;
		all other scaffolds are obtained by scrambling, i.e. changing  one or more critical hysterons. This suggests that we can characterize the
		complexity of the scaffold by the minimal number of such changes.}
	
	{The scaffolds are far less numerous than the t-graphs. For example, for $n=3$, there are only 96 scaffolds in total, and sixteen scaffolds per main loop. For $n=2$ we see that $N_{S} = 2$, and so there is only a single scaffold per main loop; this reaffirms that scrambling can only occur for $n>2$. Because they are less numerous, scaffolds facilitate exploration of t-graphs.
	}
	
	\subsection{Binary trees of transitions}\label{ssec:binary_trees}
	{We now discuss how, for a given scaffold, all (avalanche) transitions starting at a given state $S^{(0)}$ and initial direction can be generated. We note that, because avalanche may visit states only once
		due to the no-loop condition, and the magnetization spans a finite range, the number of transitions starting at a given state and initial direction is finite. We formulate the following algorithm for constructing all such possible transitions:
		\begin{algorithm}\label{alg:possible_transitions}
			Let $\{k^\pm(S)\}$ be a given scaffold. For each initial state $S^{(0)}$ and each direction (up/down):
			\begin{enumerate}
				\item Initialize the queue as consisting only of the transition $S^{(0)} \rightarrow S^{(1)}$, where $S^{(1)}$ is the landing state reached by flipping $k^\pm(S^{(0)})$.
				\item Let  $S^{(0)} \rightarrow \dots \rightarrow S^{(l)}$ be the next transition in the queue.
				\begin{enumerate}
					\item If $S^{(l)} \neq (11\dots)$, attempt to extend it in the up direction by flipping $k^+(S^{(l)})$. If the resulting transition $S^{(0)} \rightarrow \dots \rightarrow S^{(l+1)}$ is not a self-loop ($S^{(l+1)} \notin \{S^{(0)}, \dots, S^{(l)}\}$), add it to the end of the queue.	
					\item Similarly, if $S^{(l)} \neq (00\dots)$, attempt to extend it in the down direction by flipping $k^-(S^{(l)})$. If the resulting transition  is not a self-loop, add it to the end of the queue.	
					\item Save the current transition $S^{(0)} \rightarrow \dots \rightarrow S^{(l)}$ as a possible transition for the chosen state and direction, and remove it from the queue.
				\end{enumerate}
				\item Repeat step (2) until the queue is empty.
			\end{enumerate}
		\end{algorithm}
		
		Algorithm \ref{alg:possible_transitions} generates $2(2^n-1)$ sets of transitions. We note that these possible transitions for each state and initial direction can be organized in a finite binary tree (Fig.~\ref{fig:binary_trees}).
		
		\begin{example}
			We consider the $n=2$ scaffold with down boundary (1, 2), and focus on the up transitions starting from $(00)$
			(Fig.~\ref{fig:binary_trees}a).
			First, we construct the $l=1$ transition 00u - note
			the scaffold stipulates that $k^+(S) = 2$, so that 00u
			corresponds to the transition $(00) \uparrow (01)$.
			We now check if the $l=2$ extensions of this transition, 00uu and 00ud, are allowed.
			Using the scaffold, we find that 00uu  corresponds to the valid transition $(00)\uparrow(01)\uparrow(11)$, whereas 00ud leads to a
			self-loop and is forbidden; we terminate this branch
			(Fig.~\ref{fig:binary_trees}a).
			Extending 00uu, we find that 00uuu (not drawn) {does not exist} as there are no hysterons to flip up, and 00uud leads to a self-loop. All branches have now terminated,
			so we find that the only valid up sequences for transition 00U are 00u and 00uu.
			Repeating this procedure for all states and initial directions, we can construct six binary trees
			for this example scaffold {containing twelve {potential} transitions in total}(Fig.~\ref{fig:binary_trees}a).
			For $n=2$ there are only two scaffolds, and for the second scaffold we can construct {fourteen} potential transitions
			(Fig.~\ref{fig:binary_trees}b).
		\end{example}
		
		\subsection{Constructing all t-graphs}\label{ssec:constructgraphs}
		
		Given the $2(2^n-1)$ sets of possible transitions, one can construct a 'candidate' graph by selecting one of the transitions for each initial state and direction; if
		the corresponding design inequalities are consistent,
		such a candidate graph is a valid t-graph
		(section~\ref{ssec:existence}).
		By iterating over all scaffolds and selecting
		(avalanche) transitions from the corresponding binary trees, one can thus systematically generate all valid t-graphs. Formally, our algorithm is as follows:
		
		\begin{algorithm}
			Let $n$ be a given number of hysterons.
			\begin{enumerate}
				\item Generate all scaffolds for $n$ hysterons (section \ref{ssec:construct_scaffolds}).
				\item For each scaffold generated in (1):
				\begin{enumerate}
					\item Construct all $2(2^n-1)$ sets of possible transitions (Algorithm \ref{alg:possible_transitions}).
					\item Generate all candidate graphs by constructing all possible combinations of transitions from the $2(2^n-1)$ sets.
					\item Check for each candidate graph whether the design inequalities are valid (section~\ref{ssec:existence}); if so, add it to the list of valid t-graphs.
				\end{enumerate}
			\end{enumerate}
		\end{algorithm}
		
		For the $n=2$ case, by multiplying the sizes of all binary trees of transitions, we find that the first scaffold yields
		$2\times 1\times 3\times 1\times 3\times 2=36$  candidate graphs (Fig.~\ref{fig:binary_trees}a), while the second
		scaffold yields $81$ candidate graphs (Fig.~\ref{fig:binary_trees}{b}),
		yielding a total of 117 $n=2$ candidate graphs. We observe that the list of candidate graphs is often dominated by a single or small set of scaffold(s), due to the combinatorial explosion associated with large trees.
		
		To determine all valid t-graphs, we simply check for each candidate graph whether it is realizable
		(section~\ref{ssec:existence}). For $n=2$, we find that 35 of the 117 candidate graphs are realizable. {When we exclude} Garden-of-Eden states from the graph topology (section \ref{ssec:transitiongraphs}) this number reduces to only thirteen. When we ignore the intermediate states of the avalanches, the number reduces further to eleven. All of these graphs were found previously using sampling \cite{vhecke2021}; we can now state conclusively that these are the only possible graphs for $n=2$ interacting hysterons.
		
		The number of candidate graphs to be checked increases very rapidly with the number of hysterons. For just $n=3$ hysterons, we can systematically construct all trees for each of the 96 possible scaffolds, obtain all candidate graphs by multiplying the sizes of these trees for each scaffold, and then sum over the scaffolds; doing so, we obtain 4725217377852 candidate graphs. All of these candidate graphs need to be {individually} checked for realizability; thus, it is not feasible to use this method to find all {realizable t-}graphs for more than two interacting hysterons.
		
		However, the enumeration of scaffolds and trees allows to gain insight into the space of possible t-graphs. For example, the 96 scaffolds for $n=3$
		produce exactly 96 avalanche-free t-graphs.
		When GoE states are excluded, this number even reduces further to 35.
		We note in passing that a total order of the state switching fields can easily be constructed {such that an avalanche-free t-graph is obtained, for any choice of the scaffold}. For example, one
		can use a 'staircase' construction, where one
		first orders the state switching fields according to magnetization as
		$U^-(m=1) < U^+(m=0) < U^-(m=2) <U^+(m=1) < U^-(m=3) \dots$, and then arbitrarily choose an order of the state switching fields at each magnetization to obtain a well defined total order. {By substituting these state switching fields $U^\pm(S)$ for the appropriate hysteron switching fields $U_i^\pm(S)$, one can obtain a total order that produces an avalanche-free t-graph for any scaffold. Thus, all 96 avalanche-free candidate graphs for $n=3$ (or 35 when excluding GoE states) are valid t-graphs.}
		
		{We now illustrate how one can use the binary tree structure to include avalanches step by step. Considering again the $n=2$ transitions in Fig. \ref{fig:binary_trees}, we first focus on only the $l=2$ avalanches. We see that there are four of these $l=2$ avalanches per scaffold. For the scaffold with main loop (1, 2), the corresponding sequences are 00uu, 10ud, 10du and 11dd - the sequences 00ud, 01du are forbidden because of self-loops. Similarly, the scaffold with main loop (2, 1) allows for four $l=2$ transitions 00uu, 01ud, 10du and 11dd. Thus, in total, we can construct eight candidate graphs with only a single $l=2$ avalanche. We find that all these candidate graphs are realizable. }

		{We can apply the same approach for $n=3$. Considering the binary trees of transitions for each of the 96 scaffolds, we find that each of the scaffolds allows for either fourteen, sixteen or eighteen possible $l=2$ transitions - as we saw for $n=2$, this number depends on the number of $l=2$ transitions that lead to self-loops, and because these transitions must come in pairs, the total number of $l=2$ transitions is even. Summing over all scaffolds, we find that there are 1440 candidate graphs with a single $l=2$ transition. Again checking for each of these candidate graphs whether the design inequalities are consistent, we find that all these graphs are realizable.} {We believe that this phenomenon - where {\em all} candidate graphs with a single $l=2$ avalanche are realizable - extends to general $n$. To see that all such graphs are realizable, we need to construct a consistent order for the switching fields.
			To do so, one starts from a 'staircase' construction to create a total order for the t-graph without avalanches, and then only modifies the switching fields responsible for the avalanche, leading to a minor change in the ordering.}
		
		{For the next step in complexity, we have two options: we can either consider t-graphs with a single $l=3$ avalanche, or t-graphs which have two $l=2$ avalanches. We note that whereas candidate graphs with a single $l=2$ avalanche appear to always be realizable, neither candidate graphs with a single $l=3$ avalanche nor those with two $l=2$ avalanches are necessarily realizable, as can be seen from the example subgraphs in Figure~\ref{fig:impossible_subgraphs}.}
		
		{We first consider the case of a single $l=3$ avalanche, once again starting with the simple $n=2$ case shown in Fig. \ref{fig:binary_trees}. Like in the $l=2$ case, we can simply count the number of $l=3$ transitions in Fig. \ref{fig:binary_trees} to find that the scaffold with main loop (1, 2) has two possible $l=3$ transitions (10udd and 10duu) and the scaffold with main loop (2, 1) has four (00uud, 01udd, 10duu and 11ddu). Thus, one can construct six candidate graphs with a single $l=3$ avalanche. Checking the design inequalities, we find that two of these are realizable. Similarly, for $n=3$ there are 1488 possible $l=3$ avalanches and thus 1488 candidate graphs with a single $l=3$ avalanche. We find that 672 of these candidate graphs are realizable.}
		
		{Constructing and counting all graphs with two $l=2$ avalanches is slightly more demanding: in  Fig. \ref{fig:binary_trees}, we must now find all possible pairs of $l=2$ transitions. Let us first consider the scaffold with main loop (1, 2). We note that there are four combinations of state and initial direction that allow for a $l=2$ transition (00U, 10U, 01D and 11D), out of which we must choose two: the number of ways in which this can be done is $\binom{4}{2} = 6$. Furthermore, each of the initial states and directions 00U, 10U, 01D and 11D has only a single possible $l=2$ avalanche, namely 00uu, 10ud, 01du and 11dd respectively. Thus, for the $n=2$ scaffold with main loop (1, 2), there are six possible pairs of $l=2$ avalanches. The scaffold with main loop (2, 1) gives another six pairs, so that there are a total of twelve candidate graphs that have two $l=2$ avalanches. Once again checking whether the design inequalities are consistent, we find that eight of these are realizable. For $n=3$, using the same methodology, we find 9864 candidate graphs with two $l=2$ avalanches, and checking whether the design inequalities are consistent, we find that 9000 of these are realizable.}
		
		Generally speaking we see that, when even a few avalanches are included, the number of realizable t-graphs mushrooms, eventually becoming intractable. However,
		as each step of an avalanche includes additional design inequalities, we expect that more avalanche steps lead to less total orders consistent with the design inequalities, and hence a smaller volume in design space (see section \ref{sec:statweight})  -- such t-graphs, though numerous, are thus statistically rare \cite{vhecke2021}. Moreover, many of the interesting features of t-graphs rely on scrambling and the scaffold structure, rather than avalanches, and our method of separating these thus gives practical tools to explore the space of t-graphs.
		
		{We note that when the switching fields are restricted by specific physical considerations, like pairwise interactions,
			avalanches or combinations thereof might be forbidden from the outset, thus greatly reducing the number of candidate graphs. A striking example is provided by serially coupled hysterons~\cite{liu2024}, where the only possible scaffolds are Preisach scaffolds, and the only possible avalanches are $l=2$ antiferromagnetic avalanches (an up flip followed by a down flip, or vice versa). Furthermore, avalanches involving the same pair of hysterons $i, j$ are grouped together. As a result of these constraints, for three serially coupled hysterons there are only 45 candidate graphs, out of which 44 are realizable \cite{liu2024}.}
		
		\section{Statistical weight of t-graphs}\label{sec:statweight}
		\begin{figure}[t]
			\includegraphics[width=.9\textwidth]{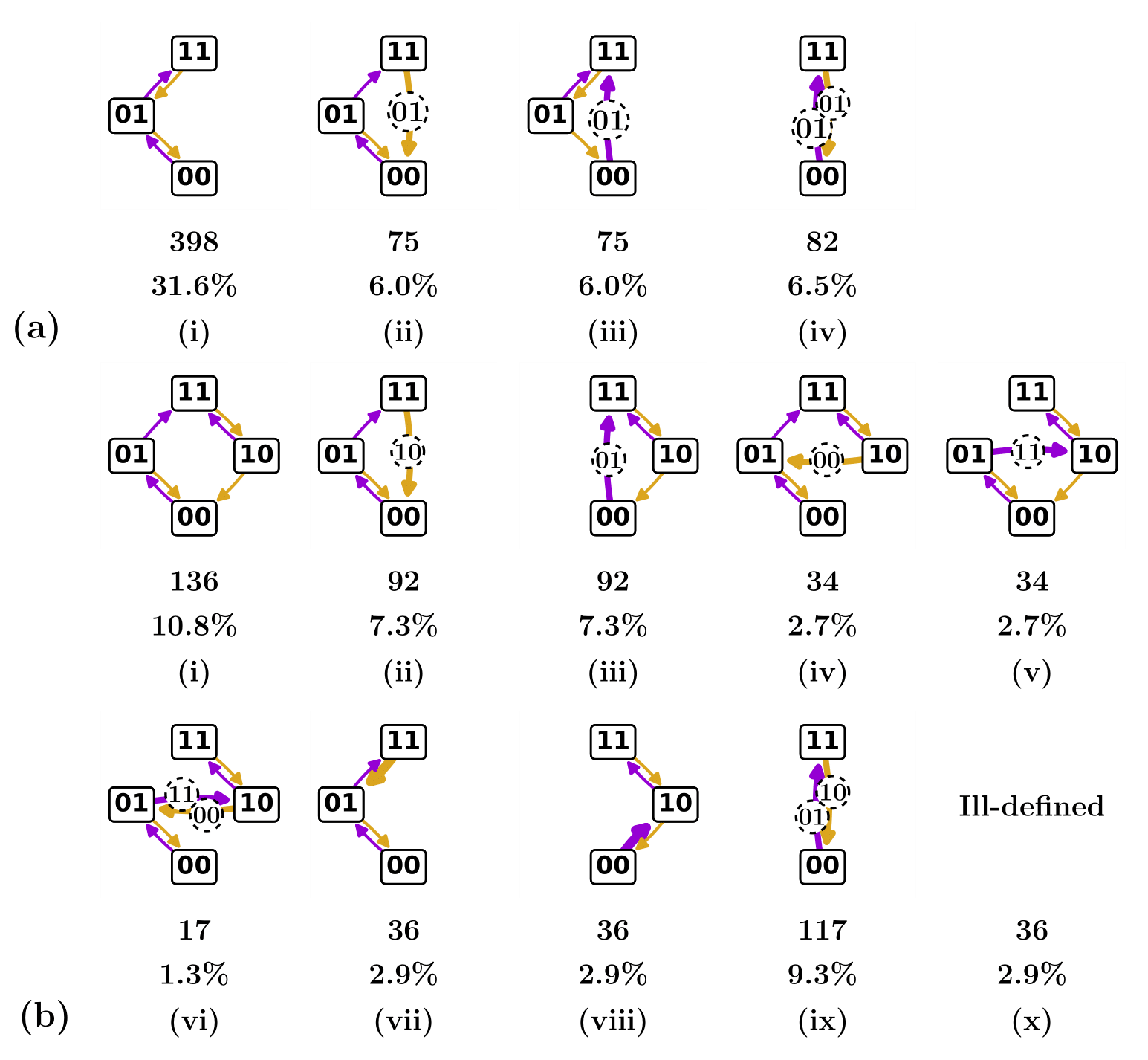}
			\caption{The thirteen $n=2$ t-graphs found by systematically constructing the candidate graphs, then checking via the design inequalities whether these are realizable; here we do not consider the GoE states. Note that, when intermediate states are ignored, graphs (a)-(ii) and (b)-(ii), and graphs (a)-(iv) and (b)-(ix) are equal, leading to the same eleven t-graphs as found previously via numerical sampling~\cite{vhecke2021}. The corresponding number of total orders and associated percentages of parameter space, when eliminating simple self-loops, are shown below each graph. Note that the the sum of total order counts (1224) is higher than that obtained when GoE states are included (850). This is because transitions associated with GoE states are now allowed to be ill-defined: this includes both race conditions and self-loops occuring in a transition initiated from a GoE state (see \ref{ssec:ill}), as well as cases where the GoE state is persistently unstable (see section~\ref{sec:drive}). (a) T-graphs, total order counts and percentages for the $n=2$ scaffold with main loop (1, 2) (see also Fig.~\ref{fig:binary_trees}).
				(b) T-graphs,  total order counts and percentages for the $n=2$ scaffold with main loop (2, 1). In panels (vii) and (viii), thicker arrows indicate $l=3$ avalanches; the full transition paths, not indicated in the figure, are $(11)\downarrow (10) \downarrow  (00) \uparrow (01)$ and $(00) \uparrow (01) \uparrow (11) \downarrow (10)$ respectively.}
			\label{fig:statistics}
		\end{figure}
		{In this section, we apply our framework to gain insight into the t-graph probabilities that were found using sampling \cite{vhecke2021}. We first use the design inequalities  to quantify the statistical weight of a single t-graph (section \ref{ssec:totorders}).  We then apply this method to all $n=2$ t-graphs found in section \ref{sec:allgraphs}, to find the percentage of graphs that is ill-defined (section \ref{ssec:analysis}).}
		
		\subsection{Domain in parameter space and counting total orders}\label{ssec:totorders}
		Previously, the parameter space of switching fields has been explored through random sampling \cite{vhecke2021}. We now show how the statistical weight of a target t-graph can be quantified via the design inequalities, and corresponding partial order.
		
		To start with, for each partial order there are one or more total orders which satisfy that partial order (or 'linear extensions', see section \ref{ssec:porder}) \cite{szpilrajn1930,pruesseruskey1994}. {The problem of generating all linear extensions for a given partial order is conceptually similar to that of generating a random linear extension, and several algorithms have been proposed based on topological sorting\cite{knuth1974structured,varol1981algorithm,kalvin1983generation}. Yet, the computational time required blows up rapidly for more complex partial orders,} and the general problem {of generating all linear extensions - or even counting these -} is in fact known to be \#P-complete \cite{brightwellwinkler1991, pruesseruskey1994}. For $n=2$, however, the number of inequalities is small, and so the problem of finding all linear extensions is manageable. Using a method based on successive counting of the number of downsets\cite{peczarski2004new}, we find that for example the graph in Fig.~\ref{fig:design_ineqs}a is associated with eighteen total orders of the switching fields.
		
		As observed by Brightwell and Winkler \cite{brightwellwinkler1991}, the problem of counting the number of linear extensions is closely related to that of computing the volume of a polyhedron. Namely, since there are $(n\cdot 2^n)!$ possible permutations of the switching field order, and by symmetry each total order takes up the same volume in parameter space, the fractional volume of a t-graph is obtained by dividing its corresponding number of total orders by $(n\cdot 2^n)!$. Applying this to the example t-graph in  Fig.~\ref{fig:design_ineqs}a, one obtains a fractional volume of $18/8! \approx 4.5\times 10^{-4}$.
		
		We check this result by directly computing the volume bounded by the design inequalities. As each design inequality forms a codimension-1 hyperplane in the design space, each t-graph corresponds to an intersection of  half-spaces, which is a convex polyhedron\footnote{Credit to M. Mungan, personal communication, 2022.} \cite{grunbaum2003}. This
		polyhedron is unbounded: we can arbitrarily increase the switching fields for a given graph as long as their order remains the same.
		Following the example of Keim and Paulsen \cite{keimpaulsen2021}, we generate the vertices of this polyhedron to generate a convex hull, and compute its volume using appropriate Python packages. To ensure the volume is finite, we set $U_i^\pm(S)\in [0, 1]$ for all switching fields: under this convention the total volume of design space is 1, and the volume of the polyhedron corresponds directly to the probability of a t-graph under random sampling. Using this approach, we indeed find a volume of $4.5\times 10^{-4}$ for our example t-graph, in agreement with our exact result.
		
		\subsection{{The space of two-hysteron t-graphs}}\label{ssec:analysis}
		Our method for generating candidate graphs via scaffolds and
		trees (section \ref{sec:allgraphs}) allows to gain insight in
		the (relative) volume in design space occupied by different t-graphs. Here we focus in particular on the fraction of design space that leads to
		ill-defined t-graphs. To do so,
		we construct the design inequalities and  all corresponding total orders for a set of well-defined t-graphs
		(see sections \ref{ssec:porder} and \ref{ssec:totorders}).
		
		First off, since we are dealing with eight switching fields, the number of possible total orders equals $8!/2! = 20160$, where we divide by $2!$ to take care of the relabeling symmetry. Each scaffold takes up an equal partition in this parameter space, as can be seen from a symmetry argument: for given state $S$, no hysteron is more likely than another to be critical. Thus, for $n=2$ each scaffold corresponds to 50 percent of parameter space, or 10080 total orders, where we note that these total orders contain both well-defined and ill-defined graphs.
		
		If we include GoE states, the 35 well-defined t-graphs for $n=2$ interacting hysterons correspond to 850 total orders of the switching fields.
		Similarly, excluding GoE states, the 13 well-defined $n=2$ t-graphs
		correspond to {1977} total orders. This suggests that only a small part of parameter space yields well-defined t-graphs.
		
		We have found that the vast majority of ill-defined graphs can be attributed to simple self-loops of length two -- as discussed in section \ref{sec:drive}, in such a case a hysteron becomes unstable in both phases as in $(00)\uparrow(01)\downarrow(00)\uparrow(01)\downarrow \dots$.
		We note in passing that this situation cannot occur in the additive pairwise model, but can easily occur when all switching fields are chosen independently.\\
		We can straightforwardly prohibit such simple self-loops by enforcing
		additional pairwise inequalities: there are four of such inequalities for $n=2$, namely $U_1^+(00) > U_1^-(10)$, $U_2^+(00)> U_2^-(01)$, $U_1^+(01)>U_1^-(11)$ and $U_2^+(10)>U_2^-(11)$. We note that these inequalities are independent of each other and of the symmetry requirement $U_1^+(00)>U_2^+(00)$, and therefore, enforcing these inequalities reduces the number of possible total orders simply to $20160/2^4 = 1260$. Accordingly, the number of total orders per scaffold reduces to 630.
		
		In the case where we include GoE states,
		the well-defined graphs still correspond to 850 total orders, as the mentioned four inequalities already emerge from each t-graph's design inequalities. Thus, by elimination of the trivial self-loops from the set of total orders, the fraction of parameter space taken up by well-defined $n=2$ t-graphs becomes $850/1260 = 67.4\%$. When we exclude the GoE states, prohibiting simple self-loops via the same four inequalities yields 1224 total orders.
		{We show the individual total order counts for each of the thirteen graphs, as well as the corresponding percentages of parameter space, in Fig.~\ref{fig:statistics}}.
		The fraction of parameter space taken up by these well-defined $n=2$ t-graphs becomes $1224/1260=97.1\%$, which is qualitatively consistent
		with earlier estimates, although those concerned a specific parametrization of the switching fields \cite{vhecke2021}.
		
		\section{Conclusion, discussion and outlook}
		
		We discussed the relation between design parameters and t-graphs for the most general model for interacting hysterons. We introduced {\em scaffolds} which allow to precisely define scrambling and facilitate the systematic construction of transitions {including} avalanches.
		We presented a systematic method for obtaining the set of design inequalities for a given t-graph (or subgraph), discussed the corresponding partial order of the switching fields, and used the partial order to straightforwardly determine the realizability of t-graphs. We showed that
		the construction and organization of t-graphs can be seen as a three-step process: first,
		all scaffolds can easily be counted and generated; second, each transition in a scaffold can be selected from easily constructible finite binary trees that encode avalanches; third, the realizability of candidate graphs formed by combining scaffolds and trees can be checked using their design inequalities. As specific examples, we count all possible t-graphs for $n=2$, and when we exclude Garden-of-Eden states, we find {thirteen} distinct t-graphs; when we furthermore ignore differences between intermediate states, we find {eleven} t-graphs, consistent with an earlier estimate based on sampling the design space \cite{vhecke2021}. For $n=3$, the number of scaffolds is $96$, producing exactly 96 avalanche-free t-graphs, and when GoE states are excluded, there are 35 $n=3$ t-graphs without avalanches; including avalanches, we find more than $4.5\times10^{12}$ candidate t-graphs, of which around $1.5\times10^4$ have been found as actual t-graphs by random sampling \cite{vhecke2021}. To enter the complex design space, we show how we can count and determine the candidate graphs and realizable t-graphs for $n=2$ and $n=3$ that contain one or two $l=2$ avalanches, or one $l=3$ avalanche. We finally discuss the statistical weight of t-graphs in design space by relating it to the
		number of total orders consistent with a given partial order.
		
		We stress that the rich structure of the t-graphs and design space can be seen as generalizing
		that of the Preisach model of non-interacting hysterons \cite{preisach1935,terzimungan2020,mungan2019structure}. First, the $2n$ hysteron-dependent switching fields of the Preisach model are generalized to $n\cdot 2^n$ state-dependent switching fields. Second, while in the Preisach model the main loop determines
		all other transitions, here the scaffold can be seen as the generalization of the main loop.
		Third, while the ordering of the hysteron dependent switching field determines the different Preisach t-graph topologies, here we need to consider the orderings of the more numerous state-dependent switching fields. However, the occurrence of avalanches and ill-defined transitions has no obvious pendant in the Preisach model, and it is these that drive the combinatorial explosion and complexity of design space.
		
		In closing, we list a number of important issues for future studies.
		
		\paragraph{Importance of avalanches.--}
		Avalanches play a mixed role. On the one hand, avalanches are not required for a variety of interesting phenomena, or even can obscure their essence: scrambling, transients and multiperiodic responses under cyclic driving do not require avalanches to occur
		\cite{deutsch2003subharmonics, keimpaulsen2021, bense2021,szulcmunganregev2022cooperative, liu2024}. Moreover,
		a scaffold-centered approach can simplify the design of experimental systems that exhibit transients \cite{liu2024,bense2021} or  subharmonic loops. On the other hand, avalanches can have a significant effect, for example allowing for transients and breaking of loop-RPM on scaffolds that are consistent with the Preisach model \cite{liu2024}. This suggests that an approach that focusses on the much smaller number of scaffolds, gradually adding a few avalanches of short length, may give already a good starting point to understand the statistics and typical response of systems of interacting hysterons: while adding many avalanches of longer lengths leads to a combinatorial explosion, and an even more severe growth of the corresponding number of total orders of the switching fields (these grow as $(n\cdot 2^n)!/n!$ which already exceeds $3\times 10^{23}$ for $n=3$), the corresponding growth of the number of design inequalities suggest that such avalanche-heavy t-graphs may only cover a
		small part of design space or even not be realizable.
		Whether such an approach truly captures the broad variety and statistics
		of transition graphs and memory effects remains an open question.
		
		\paragraph{Specific parametrizations.--}
		While in this paper we consider all the switching fields to be completely independent, different parametrizations of $U_i^\pm(S)$ have been used and may be relevant in experiments. Most prominent are several variations of pairwise additive interactions of the form $U_i^\pm(S)=u_i^\pm {-}\Sigma_j c_{ij}^\pm s_j$, where $u_i^\pm$ are the bare switching fields, and the matrices $c_{ij}^\pm$ capture the interactions -- for $c_{ij}^\pm \equiv 0$, we recover the Preisach model. Several further simplifications have been made: for example, one can assume that $c_{ij}^+=c_{ij}^-$; in addition one may assume reciprocity ($c_{ij}=c_{ji}$). {However, both non-reciprocity and $c_{ij}^+
			\neq c_{ij}^-$ can be observed in experiments \cite{bense2021,keimpaulsen2021, shohat2024geometric}. Conversely,} specific experimental settings may require even more restricted interactions, such as $c_{ij}=-d_j$, where $d_j$ are positive, for serially coupled mechanical hysterons \cite{liu2024}. Such explicit parameterizations do not affect the structure of scaffolds, trees and candidate graphs, but significantly impact
		the design {inequalities}, either by augmentation of the design inequalities with additional constraints, or by explicit conversion of the design inequalities to the specific design parameters (such as $u_i^\pm$ and $c_{ij}^\pm$). Hence, specific parametrizations lead to stricter realizability conditions and a smaller group of realizable t-graphs. In some cases, specific parametrizations may even lead to qualitative restrictions on the realizable t-graphs: for example, ferromagnetic interactions ($c_{ij}^\pm>0$) do not allow to break (loop)-RPM, and thus cannot produce transients or subharmonic orbits \cite{sethna,deutsch}; serial coupling ($c_{ij}=-d_j$) cannot produce scrambling, but can lead to breaking of (loop)-RPM via the formation of avalanches \cite{liu2024}. Gaining better insight on the relation between specific classes of interactions and t-graph topologies is an important topic for further study.
		
		In the additive pairwise coupling model and its variants, requirements on the hysteron switching fields can be formulated in terms of coupling $c_{ij}^\pm$ between hysterons, and of the 'span' of  a single hysteron, $\sigma_i = u_i^+ - u_i^-$.
		The hysteron span and coupling are identifiable in the design inequalities, even in the general model. For example, reconsidering the design inequalities in Fig.~\ref{fig:design_ineqs}e, the inequality $U_1^+(00) > U_1^+(01)$  can be interpreted as a ferromagnetic coupling (positive $c_{ij}$), where hysteron 2 causes a downward (upward) shift in the up switching field of hysteron 1 upon flipping up (down). Similarly, the inequality $U_1^+(00) > U_1^-(10)$ is associated with hysteron 1 having a positive span. In fact, the conditions that we impose to prevent trivial self-loops (section \ref{ssec:analysis}) essentially enforce that each hysteron span is always positive. Span and coupling strength may allow for a classification of different hysteron systems. First, the ratio between the coupling coefficients $c_{ij}$ and the span $\sigma_i$ quantifies the scale-invariant {\em coupling strength}, where the limit of zero coupling strength corresponds to the Preisach model \cite{vhecke2021, keimpaulsen2021}, and the limit of zero span corresponds to a spin model \cite{keimpaulsen2021}; the presence of certain t-graphs and classes of t-graphs shows powerlaw scaling with coupling strength \cite{vhecke2021,lindeman2023competition}.
		A second relevant quantity is the {\em dispersity} in hysteron spans, as evidenced by the fact that even for the Preisach model, some t-graphs require hysterons to have differing spans while others do not~\cite{terzimungan2020,mungan2019structure}.
		An approach where the design inequalities (section~\ref{ssec:design_ineqs}) and the associated partial order (section~\ref{ssec:porder}) are formulated in terms of the hysteron span and coupling may provide additional insights.
		
		\paragraph{Extended models.---} This work focussed on abstract hysterons with phenomenologically introduced interactions. More realistic models can give insight into the physical reality of hysterons, as well as the shortcomings of hysteron models.
		In addition, they may help to establish a physical picture for the interactions, and may allow to access additional physical effects. First, for hysterons where each phase is associated with a different relation between two conjugated variables - such as force and deformation - one can explicitly work out the interactions that are mediated in networks of such hysterons
		\cite{liu2024, shohat2021}. Such enhanced models are one step in a hierarchy of increasingly realistic models, that may give insight into the mechanisms that govern hysteron interactions, as well as providing design strategies to realize metamaterials that leverage such interactions. An interesting question is if we can define enhanced hysteron models that avoid race conditions and infinite loops; conversely, it is unclear
		under which conditions complex energy landscapes can still be meaningfully described by interacting hysterons.
		Second, while (thermal) noise may lead to enhanced or suppressed memories \cite{keim2011generic}, its role for interacting hysteron models is an important topic for future study. Third, many complex systems
		exhibit slow relaxations - determining which aspects
		stem from the complex transients exhibited by interacting hysterons, and which
		are due to slow relaxations of non-hysteron degrees of freedom remains an open question. Finally, this work, while general, focussed on the case of a few hysterons.
		While the continuum limit of the Preisach model has been studied in detail \cite{mayergoyz1985hysteresis,ortin1992preisach}, we have no continuum model for describing the statistics of large numbers of interacting hysterons.

		\paragraph{Total order and finite state machines.---}
		The same t-graph {\em topology} can correspond to many total orders of the switching fields (section \ref{ssec:totorders}). However, these total orders can lead to different responses when the system is subjected to specific driving protocols, as can be seen
		by considering subharmonic loops under cyclical driving \cite{keimpaulsen2021,vhecke2021}, and breaking of return point memory under asymmetric driving \cite{lindemankeim2023}. In other words, while t-graphs describe the response to arbitrary driving, extracting qualitative information, such as whether there is a cyclical driving protocol
		that produces a subharmonic orbit is not easy \cite{liu2024}.
		One strategy to effectively describe the response of systems characterized by t-graphs to specific driving inputs, such as cyclic driving or sequences of driving pulses, is to use finite state machines (FSMs) \cite{liu2024}. In brief, the idea is that any finite t-graph has only a finite number of relevant switching fields, such that the number of, e.g., pulses in $U$ that result in a different response is also finite. Defining a finite set of symbols $\{a_i\}$ then allows to determine a transition table $S'=a_i(S_j)$ that maps an initial state $S_j$ and driving $a_i$ to a final state $S'$, and therefore defines a FSM. This framework is effective in identifying specific types of responses, such as transients, for arbitrary t-graphs \cite{liu2024}.
		Moreover, this approach facilitates engineering applications, such as counting \cite{kwakernaak2023} and smart actuation for soft robots under a single input\cite{holmes2006dynamics, overvelde2015amplifying,liu2024,melancon2022}.
		
		The FSM framework is deeply linked with the total order of the state switching fields,
		with preliminary explorations indicating that any permutation in the critical switching fields produces a different transition table, and hence a different FSM.
		Hence, in addition to the design inequalities required for the t-graph topology, a
		more restricted partial order is required to ensure that the driving protocol leads the
		specific desired behavior, or to a specific FSM. Open questions for the future include how specific classes of hysteron interactions lead to, or limit, the associated FSMs and their computational power, and how to effectively design a (minimal) set of hysterons (and signals) that realize a target FSM \cite{liu2024,byun2024integrated}.\vskip6pt
		
		\enlargethispage{20pt}
		
		\funding{M.H.T. and M.v.H. acknowledge
			funding from European Research Council Grant ERC-101019474.}
		\ack{We gratefully acknowledge discussions with M. Mungan.}
		\dataccess{{All code associated with the analysis in this submission is publicly available at doi.org/10.5281/zenodo.16920054}}

		
		\vskip2pc
		

	\end{document}